\documentclass[aps,showkeys,preprint,superscriptaddress]{revtex4}
\usepackage[caption = false]{subfig}
\usepackage{graphicx}
\usepackage{float}
\usepackage{hyperref}
\usepackage[left]{lineno}
\usepackage{blindtext}
\usepackage{bm}
\usepackage{amsmath}
\usepackage{txfonts}
\usepackage{epstopdf}
\usepackage{natbib}

\usepackage{hyperref} 

\begin{document}
\title{

%Schwinger pair production by colliding laser pulse trains with quasiperiodic ordering

%
Electron-positron pair creation via Amplitude-Modulated Periodic and Quasiperiodic Pulse Sequences

%Vacuum Pair Production via Amplitude-Modulated Periodic and Quasiperiodic Pulse Sequences
%Pair Production in Multi-Pulse Electric Fields with Periodic and Quasiperiodic Modulation
%of Schwinger pair production by colliding laser pulses
%Effect of relative phase between the counterpropagating e-polarized laser pulses on $e^+e^-$ pair production via Schwinger mechanism
}
\author{Deepak Sah\footnote{Corresponding author.\\E-mail address: dsah129@gmail.com (Deepak).}}
\author{Manoranjan P. Singh}
\affiliation{Raja Ramanna Centre for Advanced Technology,  Indore-452013, India}
\affiliation{Homi Bhaba National Institute, Training School Complex, Anushakti Nagar, Mumbai 400094, India}
%\date{\today}
%\author{}
%\affiliation{}
%\affiliation{}

\begin{abstract}

%%%%%%%%%%%%%%%%%%%%%

We study nonperturbative electron–positron pair production driven by alternating-sign electric field pulse trains. Using a quantum kinetic approach, we analyze both the longitudinal momentum spectrum and the particle yield for pulse sequences with either strictly periodic temporal structure, in which the pulse amplitudes alternate in a regular and repeating $(E_1, E_2)$ pattern, or quasiperiodic (Fibonacci-ordered) structure, where the amplitudes follow a deterministic but aperiodic sequence generated by the Fibonacci substitution rule, exhibiting long-range order without exact repetition. For $N=12$ pulses, periodic trains generate regularly modulated spectra characteristic of multi-slit (Ramsey-type) interference, whereas Fibonacci sequences produce fragmented structures and partial momentum-space localization. Increasing the pulse number to $N=20$ further enhances these effects: periodic driving yields sharper and higher-contrast interference fringes, while quasiperiodic ordering leads to stronger localization and increasingly irregular spectral features.The particle yield exhibits a strongly nonlinear dependence on the field-strength ratio $\xi_{12} = E_{1}/E_{2}$. For weak modulation ($\xi_{12} \lesssim 1$), both temporal orderings produce nearly identical yields. For stronger fields ($\xi_{12} > 1$), a modest crossover behavior is observed, with quasiperiodic sequences yielding slightly larger values than the periodic case. Overall, temporal ordering primarily redistributes spectral weight in momentum space, while the integrated yield is governed predominantly by the effective field strength. These results establish long-range temporal ordering as an effective control parameter in multipulse Schwinger pair production and provide guidance for designing tailored pulse sequences in future high-intensity laser experiments.

\end{abstract}

%We study the Schwinger electron-positron pair production by a strong electromagnetic field of two colliding 
%e-polarized laser pulses

%with a relative phase shift $\Psi$. The spatio-temporal distribution of created pairs is very sensitive to this phase shift and to polarization of the pulses. We study this dependence in detail and demonstrate how it can be explained in terms of the underlying invariant field structure of the counterpropagating focused pulses. \end{abstract}
%\pacs{}
%Lorentz invariants, electron and positron ultrashort bunches}

\maketitle

\section{Introduction}

The quantum electrodynamic (QED) vacuum exhibits highly nonlinear properties when subjected to strong external fields.
Prominent manifestations include the Casimir effect, vacuum polarization, light-by-light scattering, and the nonperturbative creation of electron-positron pairs via the Schwinger mechanism~\cite{Casimir:1948dh,HeisenbergEuler,Schwinger}. 
The latter process, representing tunneling through the mass gap in the presence of an electric field, is one of the most striking predictions of strong-field QED and remains a central focus of theoretical and experimental efforts in high-intensity laser physics. Despite its fundamental importance, direct experimental observation of the Schwinger effect has remained elusive due to the extreme field strengths required, with the critical field \(E_{\mathrm{cr}} = m^2/e \sim 1.3 \times 10^{18}\,\mathrm{V/m}\) corresponding to intensities far beyond current laser capabilities.

Recent advances in ultraintense laser technology have renewed interest 
in probing these effects experimentally~\cite{DunneReview,MarklundReview}. Facilities such as ELI (Extreme Light Infrastructure) and other high-power laser systems are approaching intensities where nonlinear QED effects become accessible, with projected peak intensities reaching \(10^{23}\)--\(10^{24}\,\mathrm{W/cm}^2\) in the coming decade. Although the critical field strength remains beyond current capabilities, it is now well established that dynamically assisted and multipulse field configurations can significantly enhance pair production rates, effectively lowering the threshold for observable yields~\cite{Assisted1,Assisted2,Assisted3}. These developments have motivated extensive theoretical efforts aimed at identifying optimal temporal field structures capable of amplifying quantum coherence effects and bringing observable pair yields within reach of upcoming experiments.

An important conceptual insight emerging from these investigations is that multipulse electric fields can act as time-domain analogues of interferometers. In alternating-sign pulse sequences, electron-positron pair production exhibits interference patterns in momentum space that closely resemble Ramsey-type interferometry in atomic physics~\cite{RamseyReview,AkkermansDunne}. In this analogy, each pulse pair serves as a ``slit'' in time, and the accumulated quantum phases between successive tunneling events give rise to constructive or destructive interference depending on the pulse separation and field parameters. In particular, it has been shown that for periodic \(N\)-pulse sequences composed of identical pulses, the central peak of the momentum spectrum scales approximately as \(N^2\), reflecting coherent accumulation of quantum phases across the entire pulse train~\cite{AkkermansDunne}.

%%Temporal pulse shaping has therefore emerged as a promising tool for controlling both spectral structure and total particle yield in strong-field QED~\cite{PulseShaping,Kohlfurst}, with potential applications ranging from precision tests of QED to the development of novel pair sources.

Subsequent studies 
have demonstrated that pulse shaping and pulse ordering can be used to optimize 
particle yields and control spectral features~\cite{PulseShaping,Kohlfurst}. The dependence of the 
total particle number on the pulse number \(N\) and field parameters has been 
investigated for both fermionic and bosonic systems, revealing regimes of 
linear, quadratic, and saturation behavior ~\cite{LiBoson}.

However, all previous investigations of multipulse Schwinger pair production 
have been restricted to pulse trains with identical amplitudes—either constant 
pulses or alternating-sign sequences in which every pulse has the same peak field 
strength \(E_{0}\).  The more general scenario of amplitude-modulated pulse trains,
where the pulse amplitudes take values from a set $E_1$ and $E_2,$ has remained largely
unexplored.
Furthermore, the impact of temporal ordering in such sequences- particularly in the presence of deterministic aperiodic structures-has not been studied in the context of vacuum pair production.
This gap in knowledge is significant for several reasons.

%effect of aperiodic amplitude ordering—arranging these two amplitudes in sequences that are not strictly periodic—has never been investigated in the context of vacuum pair production.
 First, from a 
fundamental physics perspective, introducing amplitude modulation adds a new  degree of freedom to the interference problem. 
The phase accumulated between tunneling events depends not only on the temporal pulse separation but also 
on the field strengths themselves, through the dynamical phase 
\(\Theta(\bm{p},t)=2\int\omega(\bm{p},\tau)d\tau\) where 
\(\omega(\bm{p},t)=\sqrt{\epsilon_{\perp}^{2}+[p_3-eA(t)]^{2}}\) ~\cite{Li:2014psw}. Varying the amplitudes \(E_{1}\) and 
\(E_{2}\) therefore modifies the interference condition in a nontrivial way, 
potentially shifting resonance positions, altering peak widths, and even 
enabling or disabling constructive interference altogether.

%Second, from an experimental perspective, precise control over the peak field strength of individual laser pulses in an extended multipulse sequence remains a significant challenge. In realistic high-intensity laser systems, pulse-to-pulse amplitude fluctuations are unavoidable due to limitations in amplification stages, phase stabilization, and energy extraction efficiency. Deviations at the level of 20\%--50\% between successive pulses are entirely plausible in current and near-future facilities, particularly for extended pulse trains. The assumption of perfectly identical pulses therefore represents a strong idealization that warrants critical examination.
Second, from an experimental perspective, achieving precise control over the peak field strength of individual pulses in extended multipulse sequences remains challenging. Rather than treating pulse-to-pulse variations as uncontrolled noise, 
it is natural to consider deliberately engineered amplitude-modulated pulse trains. This provides a controlled framework to investigate how variations in field strength, together with their temporal ordering, influence quantum interference and particle production.

Third, the question of how aperiodic amplitude ordering affects quantum 
interference touches upon deeper issues in the physics of complex systems. 
Quasiperiodic sequences, such as the Fibonacci ordering, occupy an intermediate 
position between perfect periodicity and complete randomness. They possess 
deterministic long-range order without translational symmetry, resulting in 
singular continuous Fourier spectra that are neither discrete like periodic 
sequences nor continuous like random noise. Such sequences have been extensively 
studied in condensed matter physics in the context of quasicrystals ~\cite{2021Quasicrystal}, 
where they give rise to exotic electronic properties including critical states 
and fractal spectra. Their application to time-domain quantum interference in 
strong-field QED is entirely novel and may reveal analogous phenomena in the 
vacuum pair creation process.

In this work, we present the first systematic study of 
amplitude-modulated multipulse trains for vacuum pair production. We consider 
alternating-sign pulse sequences in which the amplitudes take two distinct values, 
\(E_{1}\) and \(E_{2}\), and compare two different types of temporal ordering.

%%study two fundamentally different ordering schemes:

In the periodic case, the amplitudes follow a regular repeating pattern,
 \(E_{1},E_{2},E_{1},E_{2},\ldots\), so that the sequence repeats every two pulses.

%Quasiperiodic amplitude modulation: The amplitudes follow a Fibonacci sequence generated recursively as \(\mathcal{F}_{n}=\mathcal{F}_{n-1}\mathcal{F}_{n-2}\) with \(\mathcal{F}_{1}=E_{1}\) and \(\mathcal{F}_{2}=E_{2}\). This introduces deterministic aperiodic long-range order while preserving the same local field strengths.
In the quasiperiodic case, the amplitudes are arranged according to a deterministic but aperiodic binary sequence (or Fibonacci type), defined explicitly for a finite number of pulses. In both cases, the amplitudes are drawn from the same set $\{E_1,E_2\}$, and the differences lies only in their ordering.

%periodic modulation with period \(2T\). This represents the natural extension of the identical-pulse case to include amplitude variation while preserving perfect temporal order.
%%
The relative strength of the two amplitudes is characterized by the ratio \(\xi_{12}=E_{1}/E_{2}\) is varied over the range \(\xi_{12}=0.25,0.50,0.75,1.0,1.5,2.0\).
This allows us to explore regimes from weak to strong amplitude contrast and to study how amplitude imbalance affects the interference pattern and particle production.
%
%between weak modulation, the ideal symmetric case, and strongly asymmetric configurations. Sub-unity values model partial depletion of alternating pulses, representing scenarios where every second pulse suffers from reduced intensity due to energy extraction limitations. Values \(\xi_{12}>1\) correspond to enhanced secondary pulses, which could arise from cumulative amplification effects or intentional pulse shaping. This systematic variation enables the first controlled exploration of how amplitude imbalance and temporal ordering jointly influence time-domain interference, momentum-space structure, and total pair yield.

Our analysis is based on the quantum kinetic framework ~\cite{QKE1,QKE2,DHW}, which 
provides a momentum-resolved description of pair production in time-dependent electric fields.
 Using this approach, we investigate the dependence of the longitudinal momentum spectrum and the total particle yield on the amplitude ratio \(\xi_{12}\), the pulse number \(N\) (with \(N=12\) and \(N=20\)), 
and the temporal ordering (periodic versus Fibonacci).

This work addresses several fundamental questions that have never been previously  investigated:

\begin{enumerate}
\item How does amplitude modulation affect multipulse interference patterns? 
Does introducing two distinct amplitudes \(E_{1}\) and \(E_{2}\) simply scale the 
identical-pulse results, or does it produce qualitatively new spectral features?

\item Can quasiperiodic amplitude ordering generate interference phenomena absent 
in periodic sequences? Does the Fibonacci sequence imprint its characteristic 
singular continuous spectrum onto the momentum distribution of created pairs?

\item Is there an optimal amplitude ratio for maximizing pair production? How do 
the resonant conditions depend on \(\xi_{12}\), and can quasiperiodic ordering 
outperform periodic driving in certain parameter regimes?

\item How does the distinction between periodic and quasiperiodic ordering scale 
with pulse number? Does increasing \(N\) amplify or diminish the differences 
between the two temporal structures?
\end{enumerate}

We demonstrate several novel findings:
First observation of amplitude-modulated interference 
fringes. The momentum spectra exhibit rich structure that depends sensitively on 
\(\xi_{12}\), with resonant peaks appearing at specific ratios 
(e.g., \(\xi_{12}\approx0.5,1.5,2.0\)) and destructive nulls at others 
(e.g., \(\xi_{12}\approx0.75,1.0\)). This reveals that the amplitude ratio is a 
powerful control parameter comparable to pulse separation.
First evidence of quasiperiodicity-induced spectral 
fragmentation. Fibonacci-ordered pulse trains produce fragmented, irregularly 
peaked spectra that differ qualitatively from the clean interference combs of 
periodic sequences. At large \(\xi_{12}\), this fragmentation evolves into 
pronounced momentum-space localization, with sharp peaks confined to narrow 
momentum windows—a time-domain analog of quasicrystal diffraction patterns.
 Near \(\xi_{12}\approx0.25\), where the periodic sequence lies in a deep destructive minimum, the Fibonacci sequence 
avoids complete cancellation and yields a modest but finite output. This 
demonstrates that aperiodic ordering can provide robustness precisely where 
periodic driving fails.

%\item[\textbf{(iv)}] \textbf{First demonstration of contrasting scaling laws.} Periodic sequences exhibit approximate \(N^{2}\) scaling of resonant peaks, confirming coherent multipulse enhancement. Fibonacci sequences show subquadratic growth (\(N^{0.7}\)), reflecting cumulative phase decoherence that prevents sustained constructive interference.

These findings establish that long-range temporal order is a fundamental control 
parameter in nonperturbative QED processes, enabling a qualitative switch between 
coherent resonant enhancement and momentum-space localization. The trade-off 
between efficiency (periodic sequences) and robustness (Fibonacci sequences) 
provides concrete guidance for designing optimized pulse sequences in the upcoming 
high-intensity laser experiments.

The paper is organized as follows. Section II presents the theoretical framework, 
including the electric field model for periodic and Fibonacci pulse trains and 
the quantum kinetic equations governing pair production. Section III contains our 
main results: momentum-resolved spectra for \(N=12\) and \(N=20\) pulses across 
the full range of \(\xi_{12}\) values, followed by integrated particle yields and 
scaling analysis. Section IV discusses the physical interpretation and implications 
for future experiments. Section V summarizes our conclusions.

%\section{Theory}\label{theory}
%\subsection{Methodology for calculating $e^+e^-$ pair production}\label{Schwinger_formula}
%Assuming the validity of a locally constant field approximation, the average number of created pairs per unit time and volume can be calculated using the Nikishov formula \cite{PhysRevLettBulanov,nikishov1970pair}:

%%

\section{Theoretical Framework}
\label{theory}
%%%%%%%%%

In this section, we introduce the electric field configurations and the quantum kinetic framework employed to study vacuum pair production. Our primary objective is to investigate how the temporal ordering of pulse amplitudes influences quantum interference effects and, consequently, the efficiency of particle production.
%%%%%%%%%%%%
\subsection{Electric field model}

We consider a temporally structured, alternating-sign multipulse electric field composed of a finite train of $N$ Sauter pulses with variable peak amplitudes. This setup allows us to systematically explore the role of temporal ordering while keeping the local pulse shape fixed.

\par

The pulses are symmetrically centered around the origin of time. The center of the first pulse is chosen as
\begin{equation}
t_0 = -\frac{N-1}{2}\,T ,
\label{eq:t0}
\end{equation}
such that the center of the $k$th pulse is located at
\begin{equation}
t_k = t_0 + (k-1)T , \qquad k = 1,2,\ldots,N ,
\end{equation}
where $T$ denotes the temporal separation between successive pulses.
Working in the temporal gauge $A_0(t)=0$, the four-potential is taken as
$A_\mu(t)=(0,0,0,A(t))$. 
The vector potential is given by
\begin{equation}
A(t)=\sum_{k=1}^{N}
(-1)^{k-1}\,
\frac{E_k}{\omega}\,
\tanh\!\left[\omega\left(t - t_k\right)\right],
\label{eq:vector_potential}
\end{equation}
where $\omega$ controls the pulse width and $E_k$ denotes the peak
amplitude of the $k$th pulse. The factor $(-1)^{k-1}$ ensures alternating polarity between successive pulses.
The corresponding electric field, $E(t)=-\dot{A}(t)$, takes the form
\begin{equation}
E(t)=-\sum_{k=1}^{N}
(-1)^{k-1}\,
E_k\,
\mathrm{sech}^2\!\left[\omega\left(t - t_k\right)\right].
\label{eq:electric_field}
\end{equation}
%%%
This construction yields a sequence of well-separated pulses with alternating sign, enabling the accumulation of dynamical phases between successive pulses and thereby generating quantum interference effects.
%%%%
In this work, the pulse amplitudes $E_k$ take values from the set ${E_1,E_2}$, allowing us to isolate the impact of temporal ordering while keeping the local field strengths fixed. We compare 
two distinct configurations: periodic and quasiperiodic pulse trains.
%%%

%%%
%In the present work, we investigate how different prescriptions for the
%pulse amplitudes $E_k$-- which takes values from the set ${E_1, E_2}$-- affect vacuum pair production.In particular, we compare periodic andquasiperiodic pulse trains constructed from the same set of local fieldstrengths.

\subsubsection*{Periodic pulse train}

For the periodic configuration, the amplitudes alternate regularly 
between $E_1$ and $E_2$. Introducing a binary variable $s_k \in \{0,1\}$, 
the amplitudes can be written as
\begin{equation}
E_k = (1 - s_k)E_1 + s_k E_2,
\end{equation}
with
\begin{equation}
s_k =
\begin{cases}
0, & k \ \text{odd}, \\
1, & k \ \text{even}.
\end{cases}
\end{equation}
%This results in a strictly periodic modulation of the field strength with period $2T$.

This defines a periodic sequence of amplitudes in which the pattern $(E_1, E_2)$ 
repeats every two pulses, corresponding to a temporal repetition scale of $2T$.

\subsubsection*{Quasiperiodic pulse train}

For the quasiperiodic configuration, the pulse amplitudes are assigned 
according to a deterministic aperiodic binary sequence. 
In this work, we consider finite sequences of length $N=12$ and $N=20$. A representative 
example for $N=12$ is given by
\begin{equation}
\{s_k\} = \{0,1,0,0,1,0,1,0,0,1,0,0\},
\end{equation}
with the corresponding field amplitudes defined as

%such that
\begin{equation}
E_k = (1 - s_k)E_1 + s_k E_2.
\end{equation}
This sequence represents a Fibonacci-type quasiperiodic ordering, 
introducing long-range aperiodic correlations while preserving the 
same set of local field amplitudes.

The distinction between periodic and quasiperiodic arrangements lies 
solely in the temporal ordering of the pulses. 

%Physically, this difference modifies the phase accumulation between successive pulses, leading to qualitatively different interference patterns in the time domain.

%As a consequence, one expects observable changes in both the momentum spectra and the total yield of produced pairs, reflecting the sensitivity of strong-field processes to the temporal structure of the driving field.

This setup enables a systematic comparison between periodic and 
quasiperiodic driving, allowing us to isolate the role of field-strength  modulation in shaping the momentum spectra of 
created particle--antiparticle pairs.

%\subsubsection*{Parameter specifications}

In all numerical calculations, we fix $E_2 = 0.1\,E_{\mathrm{c}}$, where 
$E_{\mathrm{c}} = m^2/e$ denotes the Schwinger critical field. 
We introduce the dimensionless field-strength ratio
\begin{equation}
\xi_{12} = \frac{E_1}{E_2},
\end{equation}
and consider the values
\begin{equation}
\xi_{12} = 0.25,\;0.50,\;0.75,\;1.0,\;1.5,\;2.0.
\end{equation}
This range spans weak modulation ($\xi_{12}=0.25$), the symmetric case 
($\xi_{12}=1.0$), and strongly asymmetric configurations ($\xi_{12}=2.0$), 
thereby enabling a systematic exploration of different interference regimes.

The pulse duration is fixed to $\tau = 1/\omega = 10\,[m^{-1}]$, and the 
interpulse separation is chosen as $T \simeq 8\tau$, ensuring well-separated 
pulses. Throughout this work, we employ natural units $\hbar = c = 1$ and 
set the electron mass to $m=1$.

%%
%The Fibonacci sequence is particularly well suited for this study because it possesses a sigular continuous Fourier spectrum-neither discrete like periodic sequences nor continuous like random noise- which is expected to manifest in the momentum distribution of created pairs.

%%

\subsection{Quantum kinetic Description}

To describe electron--positron pair production in these time-dependent fields, we employ
a quantum kinetic approach based on the quantum Vlasov equation (QVE), which naturally incorporates non-markovian effects and quantum interference through dynamical phase accumulation 
 \cite{Banerjee:2018azr,Panferov:2015yda,Otto:2018jbs}.
%%%%%%%%%%%%%%%%%%%%%%%%

We restrict our analysis to the subcritical field,
$E < E_{\mathrm{c}}$, where pair production and the associated
backreaction current remain small. 
This allows us to neglect collision
effects as well as the feedback of the produced particles on the
background field.Furthermore, the spatial scale of typical laser
pulses is much larger than the electron Compton wavelength,
justifying the neglect of spatial inhomogeneities. In the case of
counterpropagating pulses forming  a standing wave,  magnetic field effects near the antinodes are negligible,
so that $B(t)\approx 0$.
\par
Starting from the Dirac equation in a homogeneous electric field and
applying a time-dependent Bogoliubov transformation, the QVE can be
formulated as an integro-differential equation governing the evolution of
the single-particle momentum distribution function $f(\bm{p},t)$:
\begin{equation}
\frac{d f(\bm{p}, t)}{dt} = \frac{\lambda(\bm{p},t)}{2}
\int_{t_0}^{t} dt' 
\lambda(\bm{p}, t')\,[1 - 2 f(\bm{p}, t')]
\cos\!\left[\Theta (\bm{p}, t, t')\right],
\label{eqn1}
\end{equation}
where
\begin{equation}
\lambda (\bm{p},t) = \frac{e E(t)\,\varepsilon_{\perp}}
{\omega^2(\bm{p}, t)}
\end{equation}
is the vacuum transition amplitude and
\begin{equation}
\Theta (\bm{p},t,t') = 2 \int_{t'}^{t} d\tau \, \omega(\bm{p}, \tau )
\end{equation}
is the dynamical phase.

%%%%%%%%%%%%%%%%%%% which governs the constructive and destructive interference of pair produciton amplitudes generated at different times.
The quasiparticle energy $\omega(\bm{p},t)$, transverse energy
$\varepsilon_{\perp}$, and longitudinal kinetic momentum $P_3(t)$ are
defined as
\begin{align}
\omega(\bm{p}, t) &= \sqrt{\varepsilon_{\perp}^2 + P_3^2(t)}, \\
\varepsilon_{\perp} &= \sqrt{m^2 + p_{\perp}^2}, \\
P_3(t) &= p_3 - eA(t),
\end{align}
where $\bm{p}=(p_{\perp},p_3)$ denotes the canonical momentum.
%%%%
In the asymptotic limit $t\to+\infty$, where the external field vanishes,
the distribution function $f(\bm{p},t)$ corresponds to the number density
of real particles created with momentum $\bm{p}$. Our analysis focuses on this asymptotic distribution.
\par
For numerical implementation, Eq.~\eqref{eqn1} is recast into an equivalent system of three coupled ordinary differential equations,
\begin{align}
\frac{df(\bm{p},t)}{dt} &= \frac{1}{2}\lambda(\bm{p},t)\,u(\bm{p},t), \\
\frac{du(\bm{p},t)}{dt} &= \lambda(\bm{p},t)\,[1-2f(\bm{p},t)]
-2\omega(\bm{p},t)\,v(\bm{p},t), \\
\frac{dv(\bm{p},t)}{dt} &= 2\omega(\bm{p},t)\,u(\bm{p},t),
\end{align}
with the initial conditions
\begin{equation}
f(\bm{p},-\infty)=u(\bm{p},-\infty)=v(\bm{p},-\infty)=0.
\end{equation}
The auxillary functions $u(\bm{p},t)$ and $v(\bm{p},t)$ 
encode  the real and imaginary parts of correlation functions and contain information about the phase coherence of the created pairs ~\cite{Sah:2025nbr,Banerjee:2018fbw}.
%%%
The number density of created particles is obtained by integrating the distribution function over momentum, 
\begin{equation}
n(t) = 2 \int \frac{d^3 p}{(2\pi)^3} \, f(\bm{p},t),
\end{equation}
where the factor of $2$ accounts for spin degeneracy.
\par
In this work, we focus on the asymptotic particle distribution  $f(\bm{p},+\infty)$ and the corresponding particle number  density $n(+\infty)$, which provide direct measures of particle production efficiency.
Furthermore, we restrict our analysis to 
vanishing transverse momentum, $p_{\perp}=0$,
where the production probability is maximal for Sauter-type pulses. This allows us to isolate the essential interference effects encoded in the longitudinal momentum dynamics.

\section{Results}

%%%We now present and discuss the momentum-resolved spectra and yields of electron-positron pairs produced by alternating-sign multi-pulse electric fields. Our analysis focuses on the role of long-range temporal order by comparing strictly periodic pulse trains with quasiperiodic Fibonacci sequences. The results are organized according to increasing the field-strength ratio $\xi_{12} = E_1/E_2,$ followed by an examination of pulse-number dependence.%%%%%%
%%%%%%
%%%%%%
%%%%%%
We present momentum spectra and yields for pair production in alternating-sign multi-pulse fields, comparing periodic and quasiperiodic sequences. Results are organized by increasing $\xi_{12}=E_1/E_2$ and pulse number.
%%%%%%
%%%%%%
%%%%%%

%%%%%%%%%

%%%%%%%%%%%%%

%%%%%%%%%%%%%%%

\begin{figure}[tbp]
\centering
{\includegraphics[width=0.9682\columnwidth]{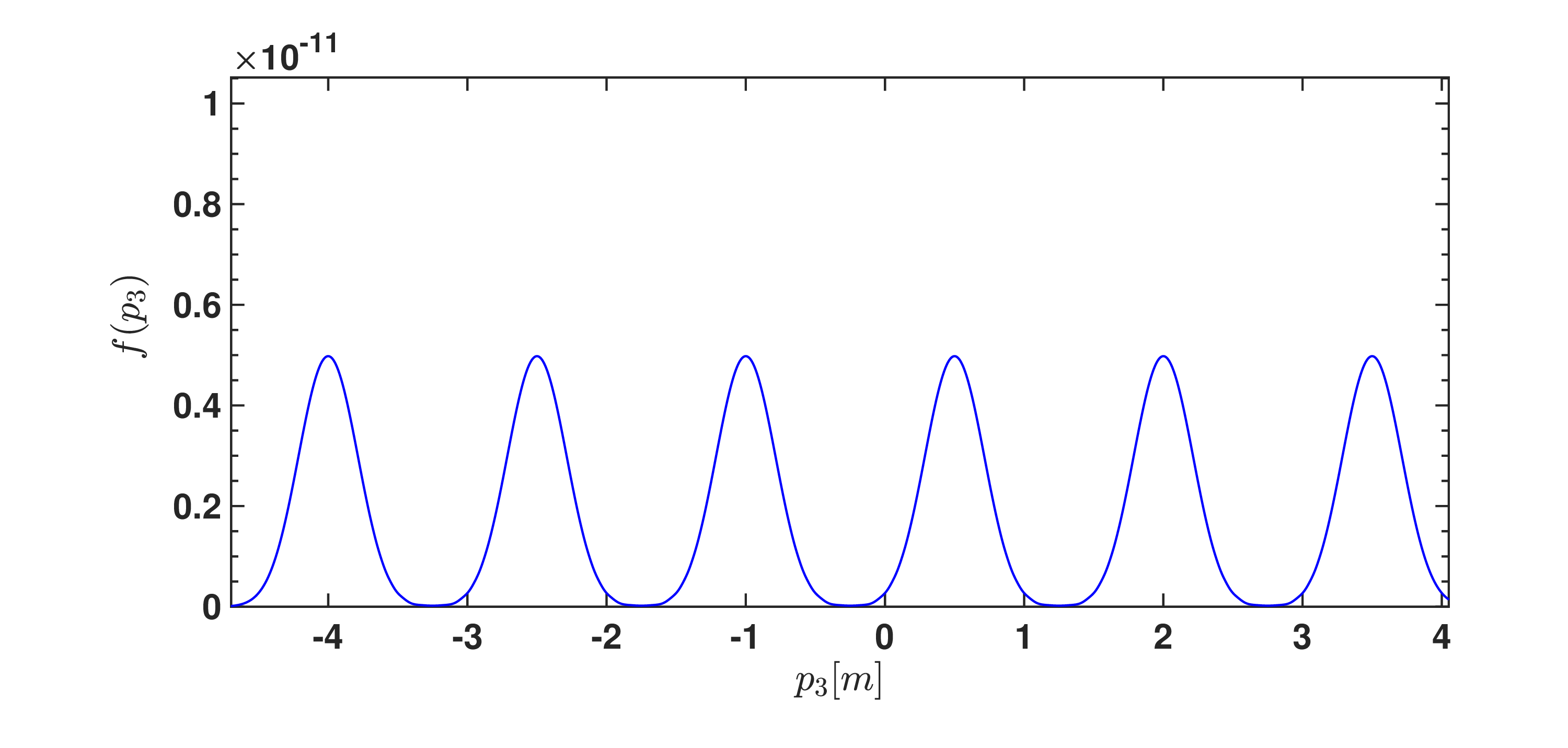}
\includegraphics[width=0.9682\columnwidth]{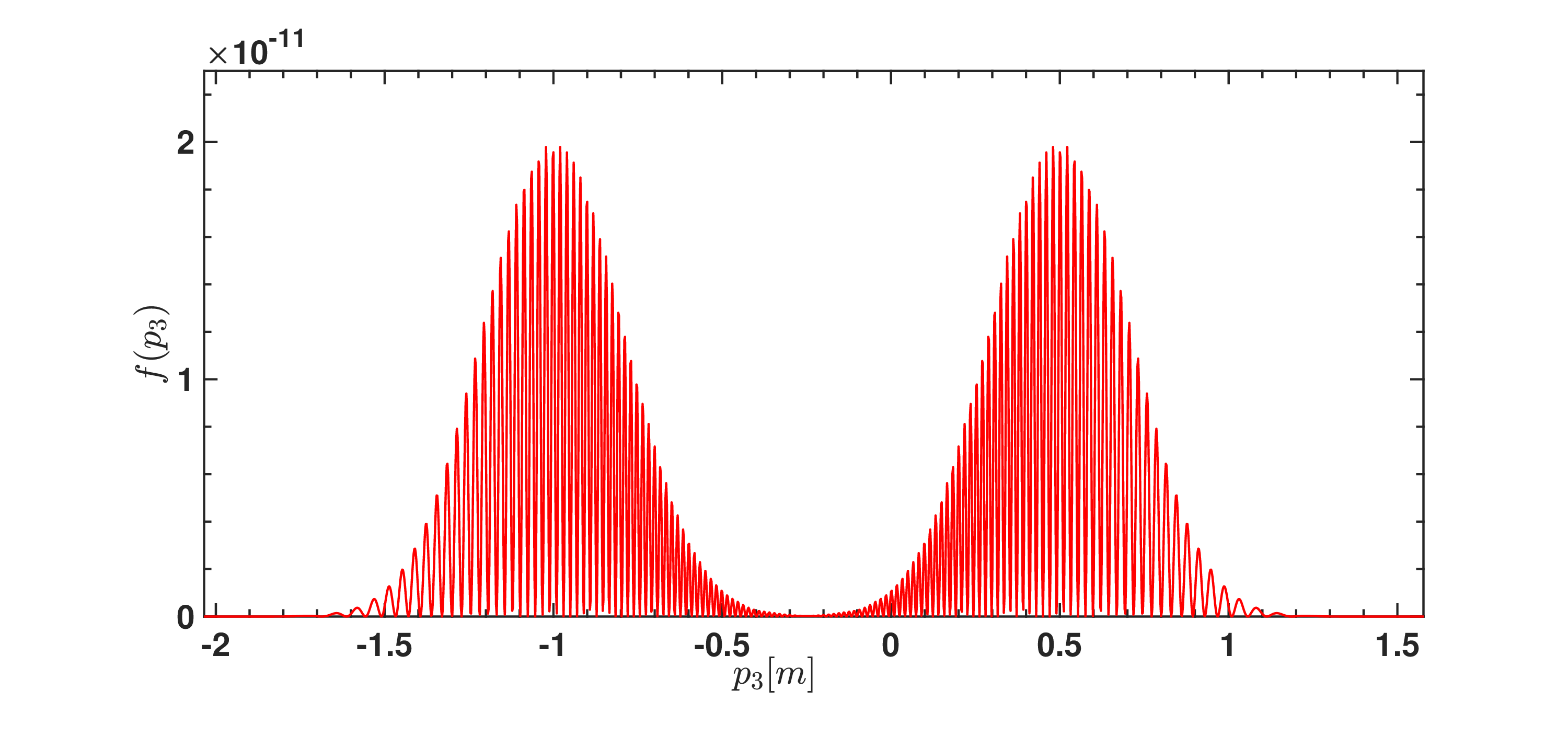}}
\caption{\label{Fig1} Longitudinal momentum spectra
$f(p_3)$  for \(N=12\) pulses at field-strength ratio \(\xi_{12}=0.25\). Upper panel: periodic sequence; lower panel: Fibonacci quasiperiodic sequence. Parameters: \(E_2=0.1E_{\mathrm{c}}\), \(\tau=10\,[m^{-1}]\), \(T=78.2\,[m^{-1}]\), \(p_\perp=0\). All quantities in electron mass units (\(m=1\)).}
\end{figure}

Figure~\ref{Fig1} shows the longitudinal momentum distribution \(f(p_3)\) for an alternating-sign train of \(N = 12\) pulses at a strongly imbalanced field-strength ratio \(\xi_{12} = 0.25\), with \(E_2 = 0.1\,E_c\) and vanishing transverse momentum.

For the periodic pulse sequence [upper panel of Fig.~\ref{Fig1}] , the spectrum exhibits a regular comb-like interference structure consisting of symmetrically spaced peaks,
reflecting coherent phase locking imposed by the strict temporal periodicity of amplitude $(E_1, E_2)$ of the pulse train. 
However, the overall magnitude remains strongly suppressed, \(f(p_3) \sim 10^{-12}\) 
due to the large amplitude imbalance $E_1 \ll E_2,$ which inhibits efficient accumulation of transition amplitude.
%%

%%%%%%%%%%%%%%%
In contrast, the quasiperiodic sequence [lower panel of Fig.~\ref{Fig1}] produces a dense set of rapidly oscillating peaks modulated by a smooth envelope, with two broad lobes centered around $p_3 \approx 0.$
The absence of a simple periodic pattern reflects the lack of long-range temporal order and the resulting breakdown of global phase matching. Nevertheless, the overall magnitude of the spectrum remains comparable to the periodic case, \(f(p_3) \sim 10^{-11}\), indicating that quasiperiodicity  of amplitude modulation does not lift the strong suppression imposed by the weak modulation.

Thus, at \(\xi_{12} = 0.25\), both configurations lie in a deeply suppressed regime: periodic ordering preserves regular interference without enhancement,
while quasiperiodicity leads to fragmented oscillations without increasing yield.

%Periodic driving preserves regular interference without enhancement, while Fibonacci ordering replaces it with fragmented oscillations, but neither configuration yields significant creation of pairs.

\begin{figure}[tbp]
\centering
{\includegraphics[width=0.9682\columnwidth]{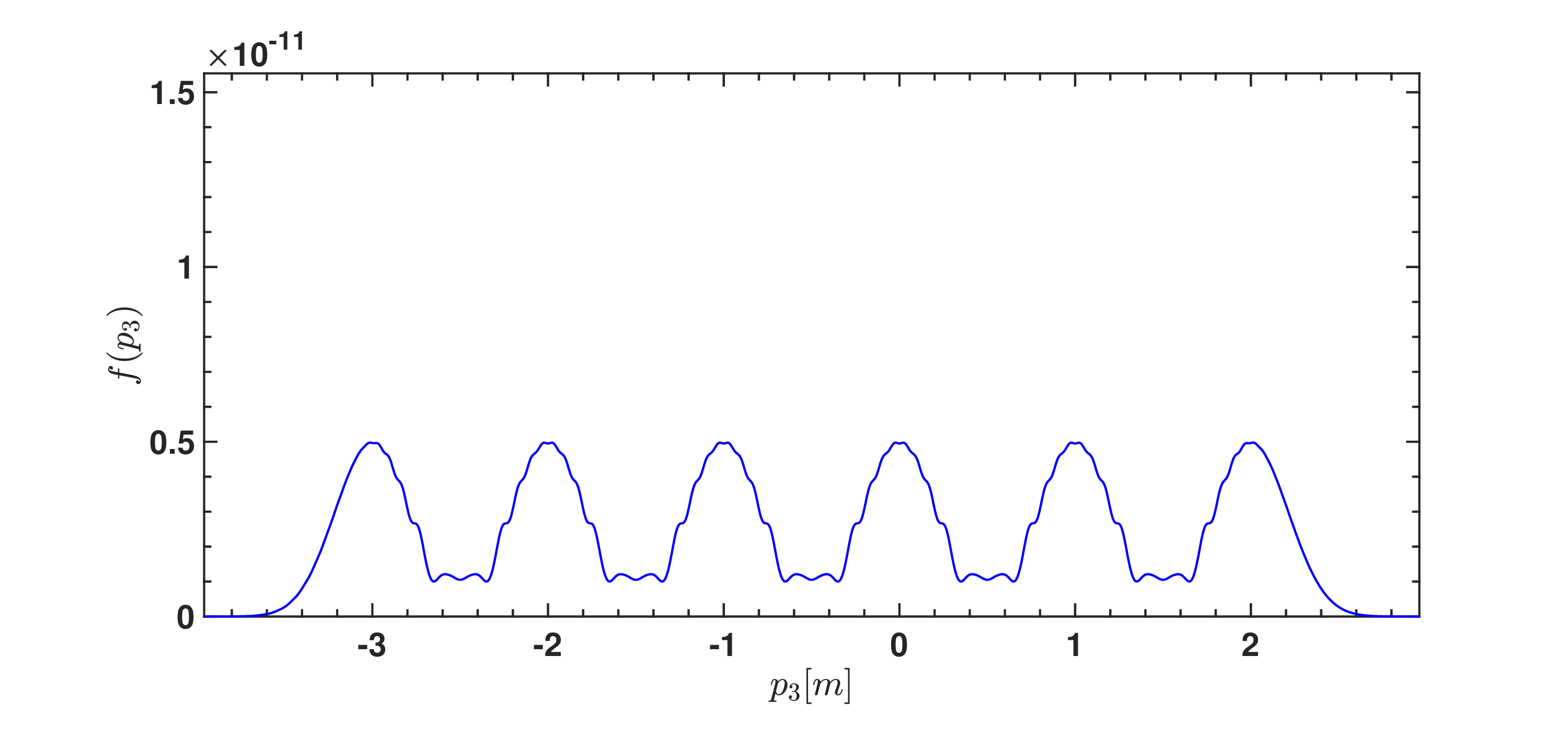}
\includegraphics[width=0.9682\columnwidth]{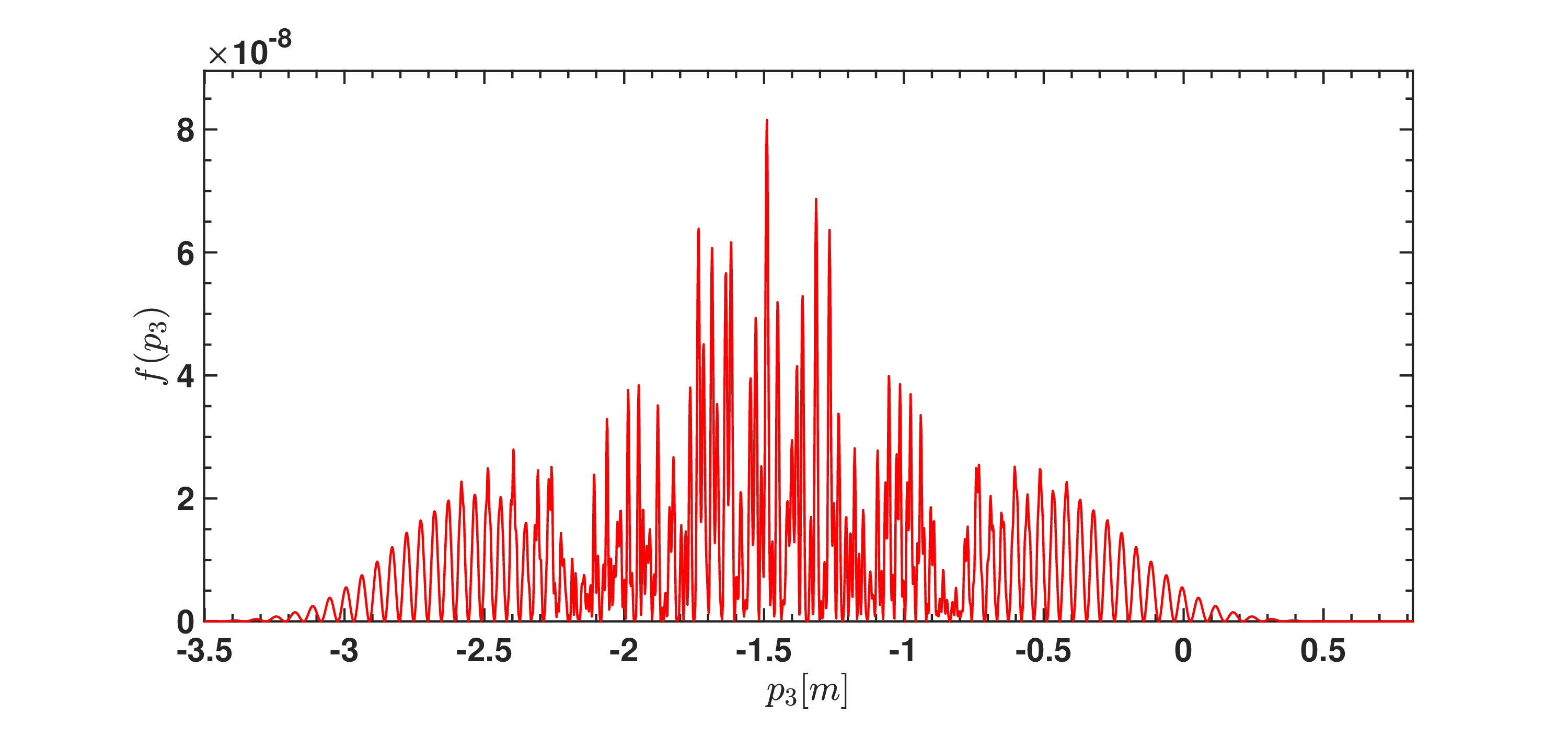}}
\caption{\label{Fig2} Same as Fig.~\ref{Fig1} but for $\xi_{12} =0.50$}
\end{figure}

Figure~\ref{Fig2}  shows the spectra at \(\xi_{12} = 0.50\), where the pulse amplitudes $(E_1,E_2)$ become comparable.
For the periodic sequence (see Fig.~\ref{Fig2}), a high-contrast interference comb emerges, with pronounced maxima and minima symmetrically distributed around $p_3 = 0$. Forming a smooth oscillatory envelope characteristic of constructive multi-pulse interference. This behavior signals efficient phase matching across the pulse train and marks the onset of resonant enhancement.
By contrast, the quasiperiodic sequence produces an irregular spectrum 
with fragmented peaks distributed over a limited momentum range. While local enhancements occur, no sharply isolated resonances emerge. 
%The quasiperiodic ordering disrupts global phase coherence, redistributing spectral weight across many momentum channels rather than concentrating it into resonant peaks.

This comparison demonstrates that, at moderate modulation depth, temporal ordering alone can switch the system between coherent resonant enhancement (periodic case) and incoherent, fragmented production (quasiperiodic case).

%%%

\begin{figure}[tbp]
\centering
{\includegraphics[width=0.9682\columnwidth]{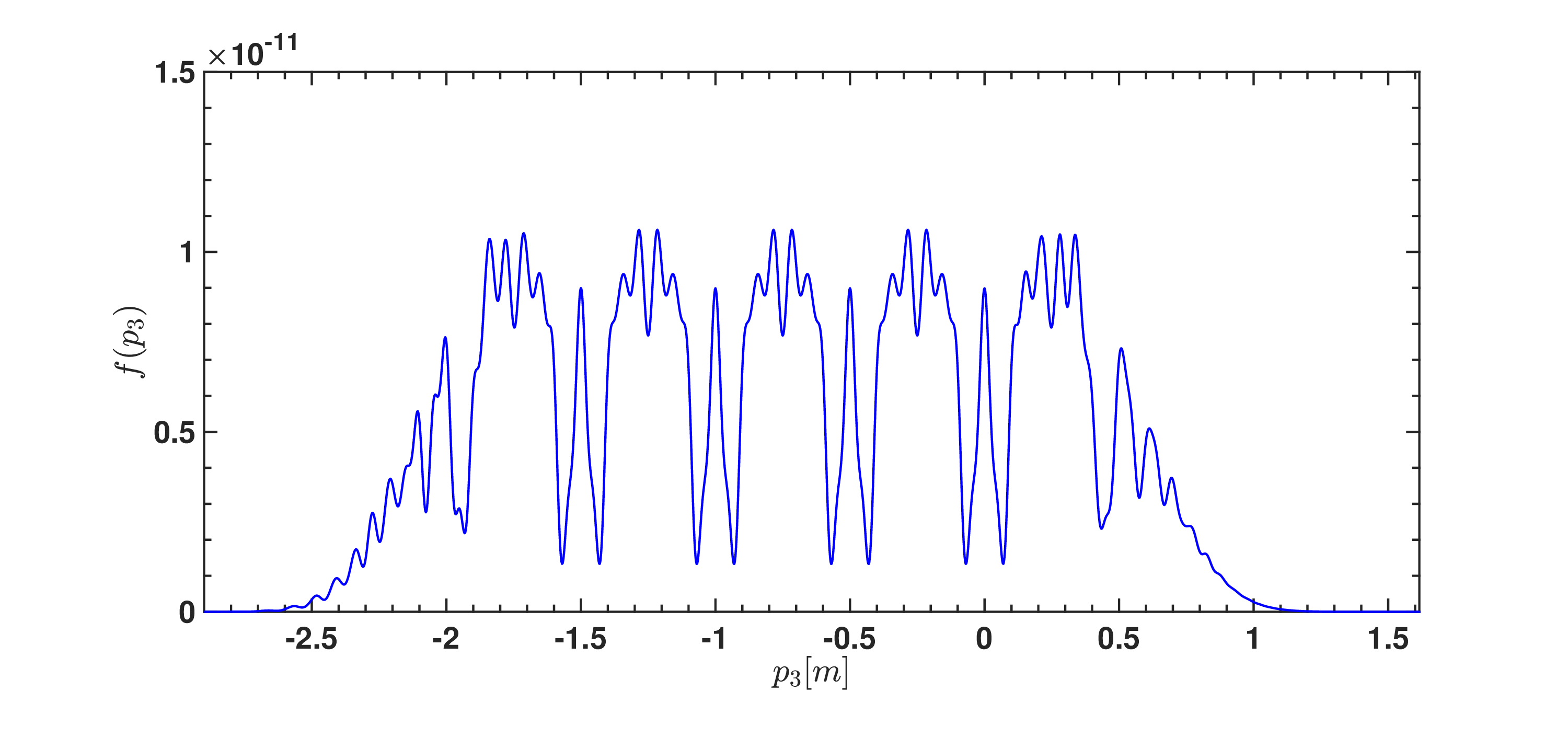}
\includegraphics[width=0.9682\columnwidth]{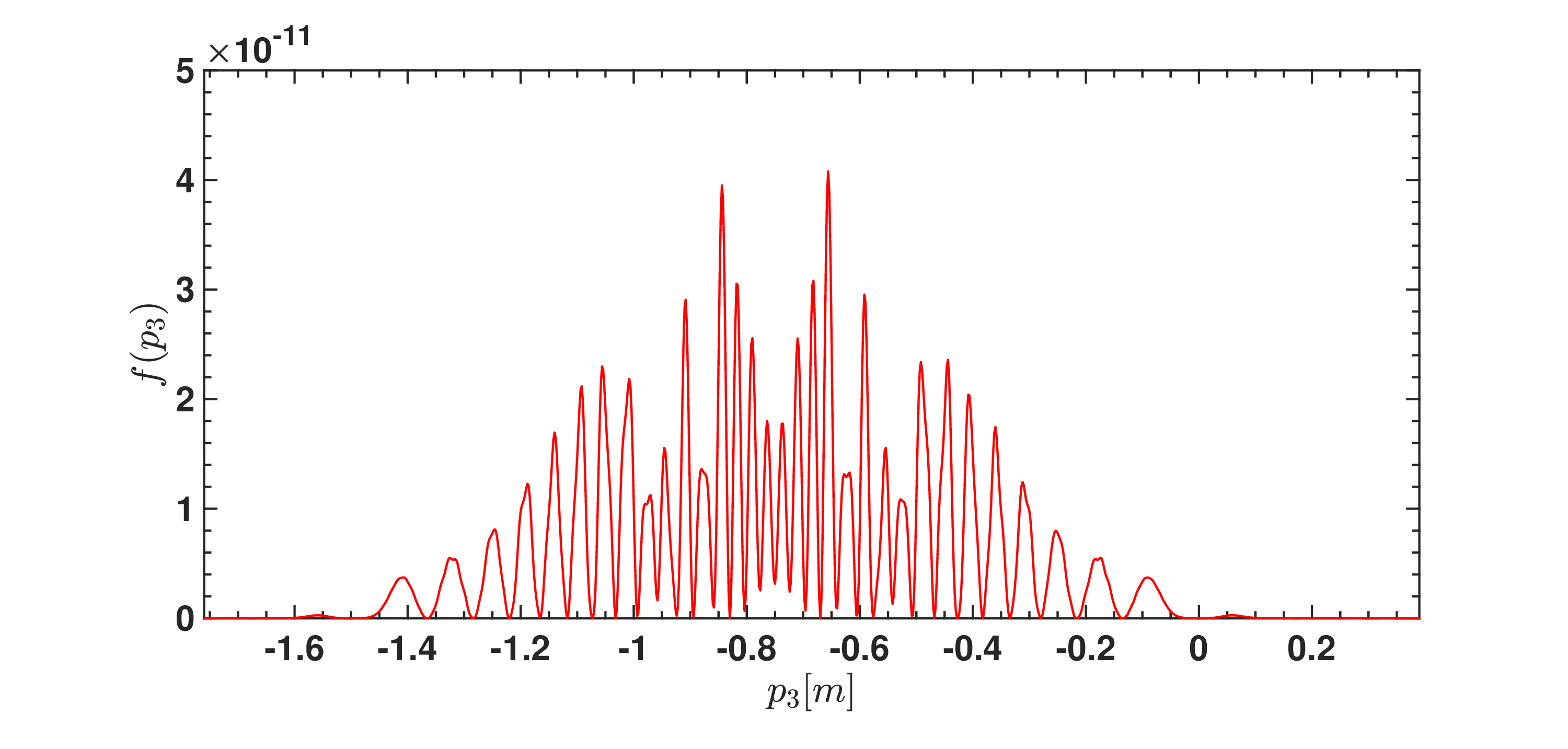}}
\caption{\label{Fig3} Same as Fig.~\ref{Fig1} but for $\xi_{12} =0.75$}
\end{figure}

%As $\xi_{12} $ increases further, the system enters a different interference regime.
%%
At \(\xi_{12} = 0.75\), shown in Fig.~\ref{Fig3},
the system enters a distinct interference regime.
%%

%the two pulse amplitudes become comparable,enabling oscillatory behavior characteristic of quantum interference in multipulse configurations.
%%

For the periodic amplitude sequence, the spectrum remains strongly suppressed despite retaining an ordered oscillatory structure. The distribution is confined within a broad envelope with peak heights limited to \(f(p_3) \sim 10^{-11}\). This behavior indicates near-perfect phase cancellation across successive pulses, corresponding to a destructive interference resonance.

In contrast, the quasiperiodic sequence avoids such cancellation, exhibiting  irregularly peaks
with locally enhanced amplitudes.
 The aperiodic temporal ordering disrupts the precise phase relations responsible for destructive interference.
 
 Thus, \(\xi_{12} = 0.75\) represents an regime where quasiperiodicity of amplitudes provides a relative enhancement by suppressing destructive  resonances present in the periodic driving.

\begin{figure}[tbp]
\centering
{\includegraphics[width=0.9682\columnwidth]{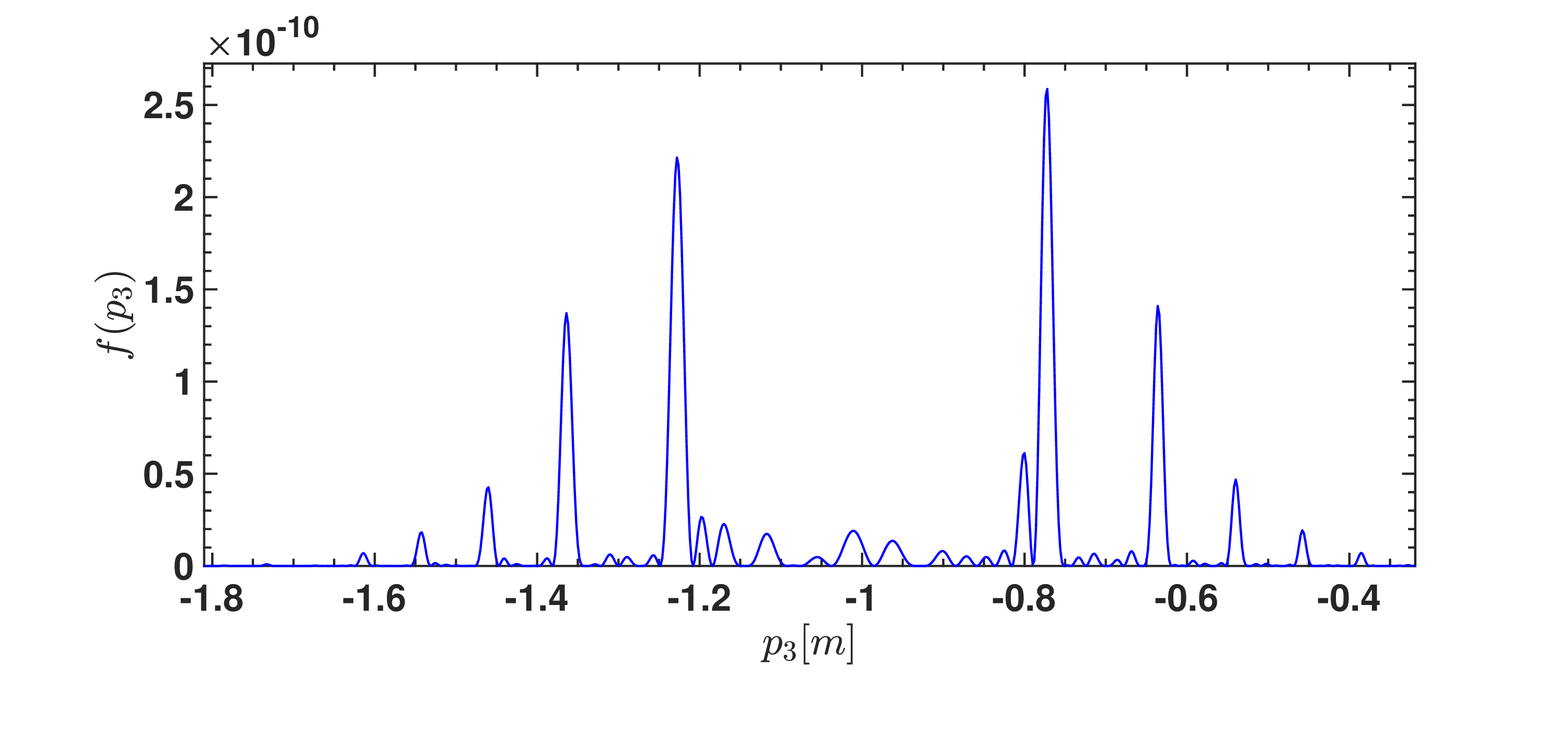}
\includegraphics[width=0.9682\columnwidth]{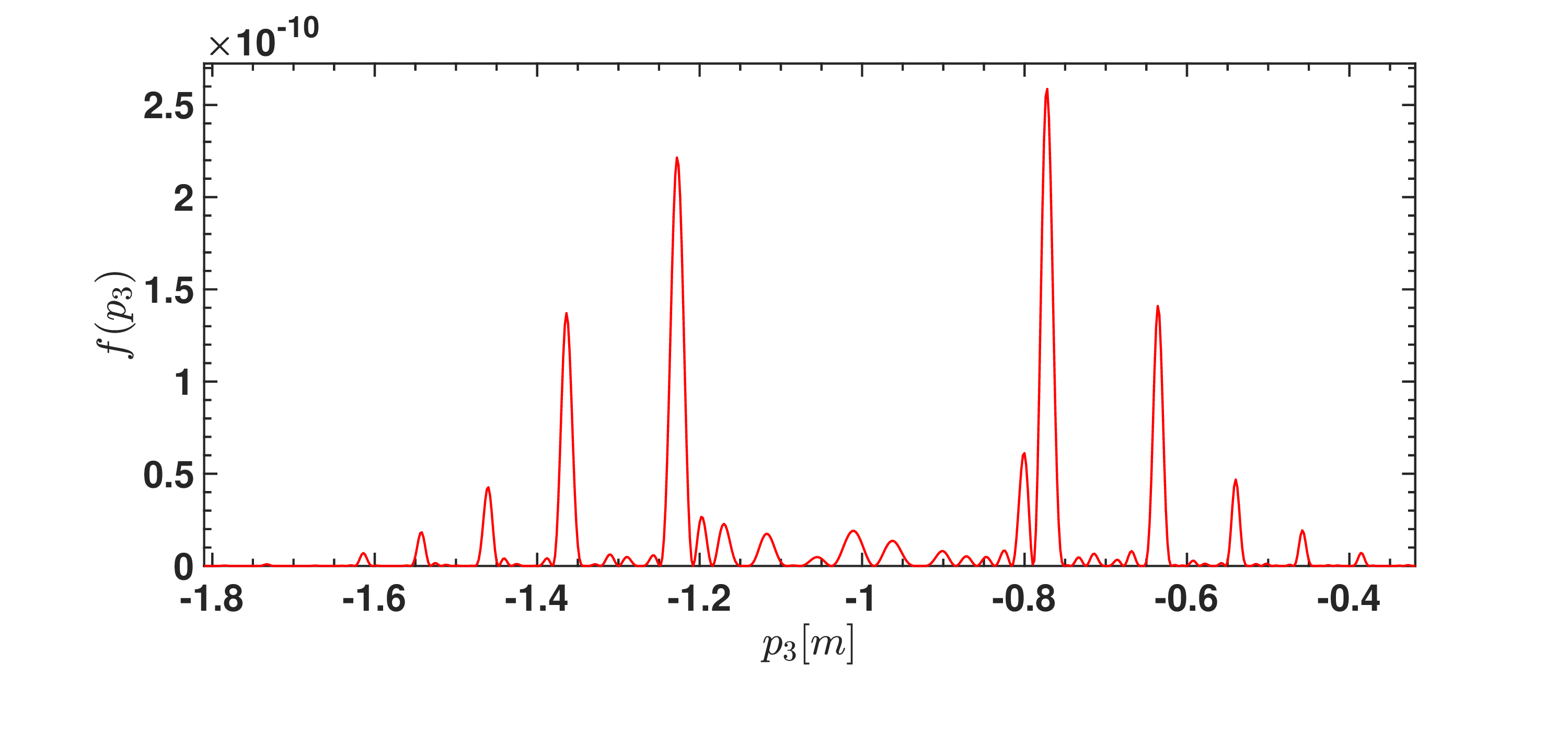}
}
\caption{\label{Fig4}Same as Fig.~\ref{Fig1} but for $\xi_{12} =1.0$}
\end{figure}

 At $\xi_{12} =1.0$ (Fig.~\ref{Fig4}), corresponding to equal amplitudes  \(E_1 = E_2\),
 the periodic sequence exhibits strong momentum selectivity, with the spectrum collapsing into a few sharply localized peaks.  This behavior reflects a symmetry-driven interference condition analogous to a temporal Ramsey interferometer. The quasiperiodic sequence shows similar localization due to the identical amplitudes.
 
 %In this symmetric configuration of amplitudes, the alternating-sign pulse train acts as a highly selective temporal  Ramsey interferometer ~\cite{AkkermansDunne}.
%%
%The spectrum collapses into a small number of sharply localized peaks, while pair production is strongly suppressed elsewhere in momentum space. Only those momenta satisfying a stringent phase-matching condition survive, leading to extreme momentum selectivity. This behavior reflects a symmetry-driven interference condition imposed by identical pulse amplitudes and alternating polarity.

%The Fibonacci spectrum exhibits similar  localization since both amplitude $E_1 =E_2.$

\begin{figure}[tbp]
\centering
{\includegraphics[width=0.9682\columnwidth]{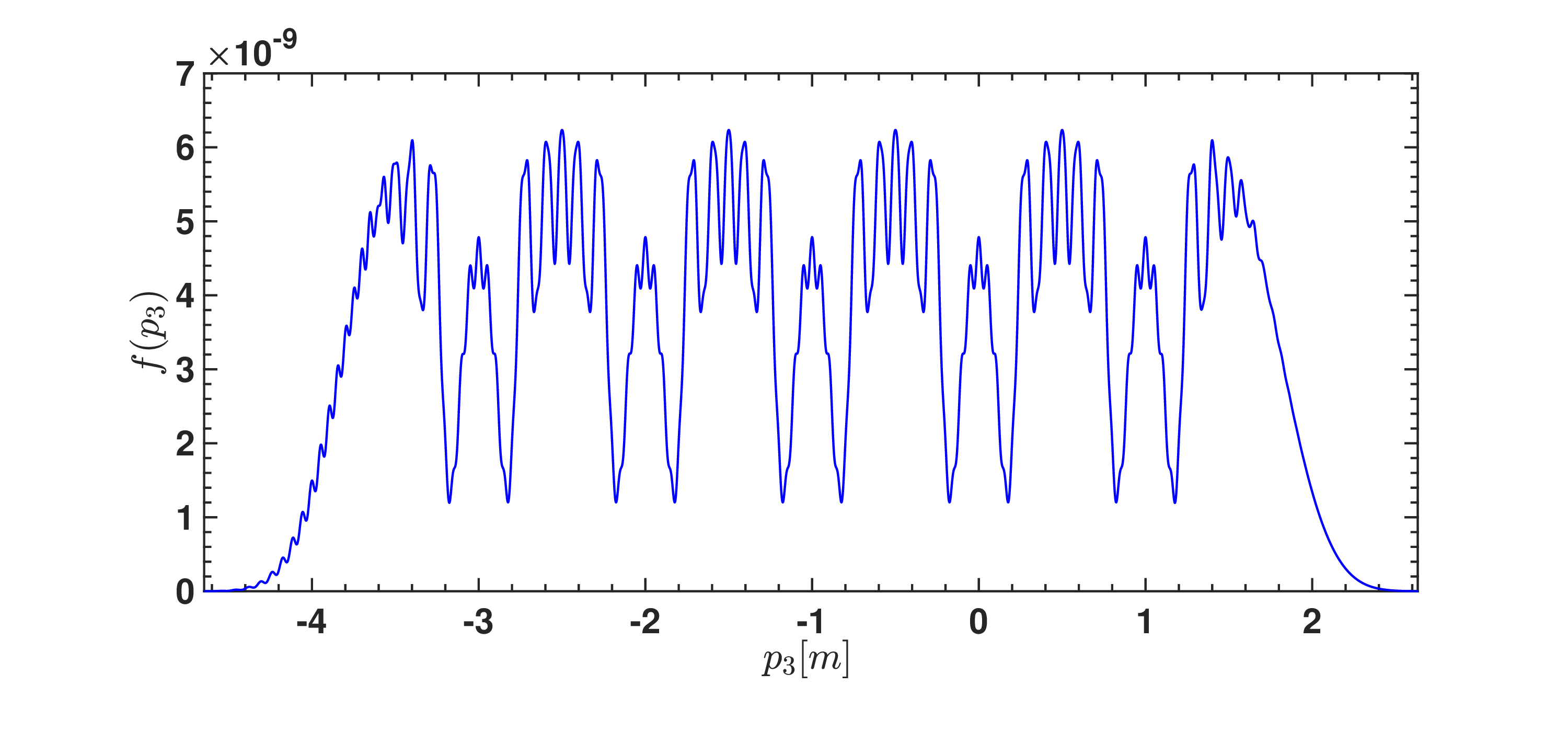}
\includegraphics[width=0.9682\columnwidth]{N_12_e12_150_fib_new.eps}}
\caption{\label{Fig5} Same as Fig.~\ref{Fig1} but for $\xi_{12} =1.50$}
\end{figure}

\begin{figure}[tbp]
\centering
{\includegraphics[width=0.9682\columnwidth]{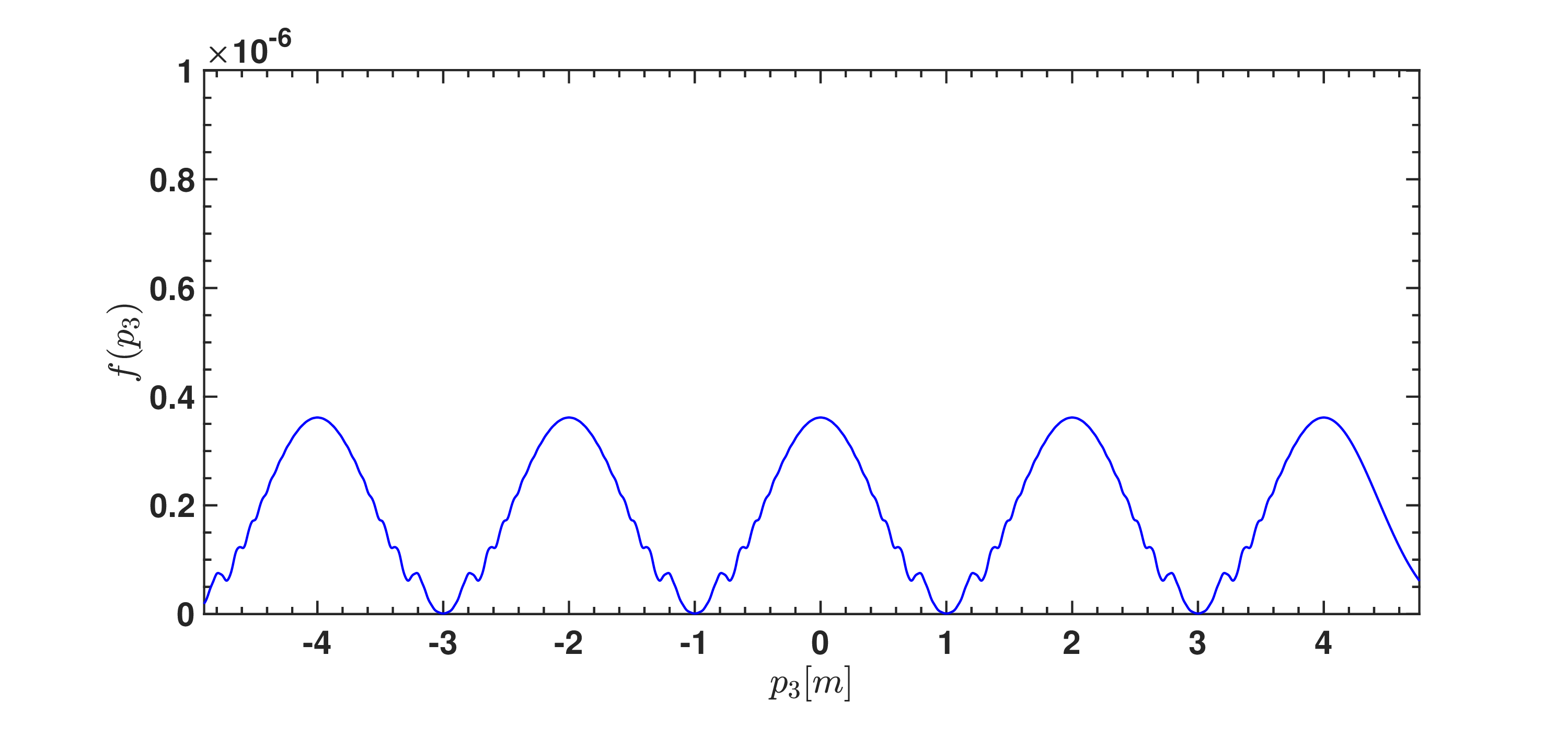}
\includegraphics[width=0.9682\columnwidth]{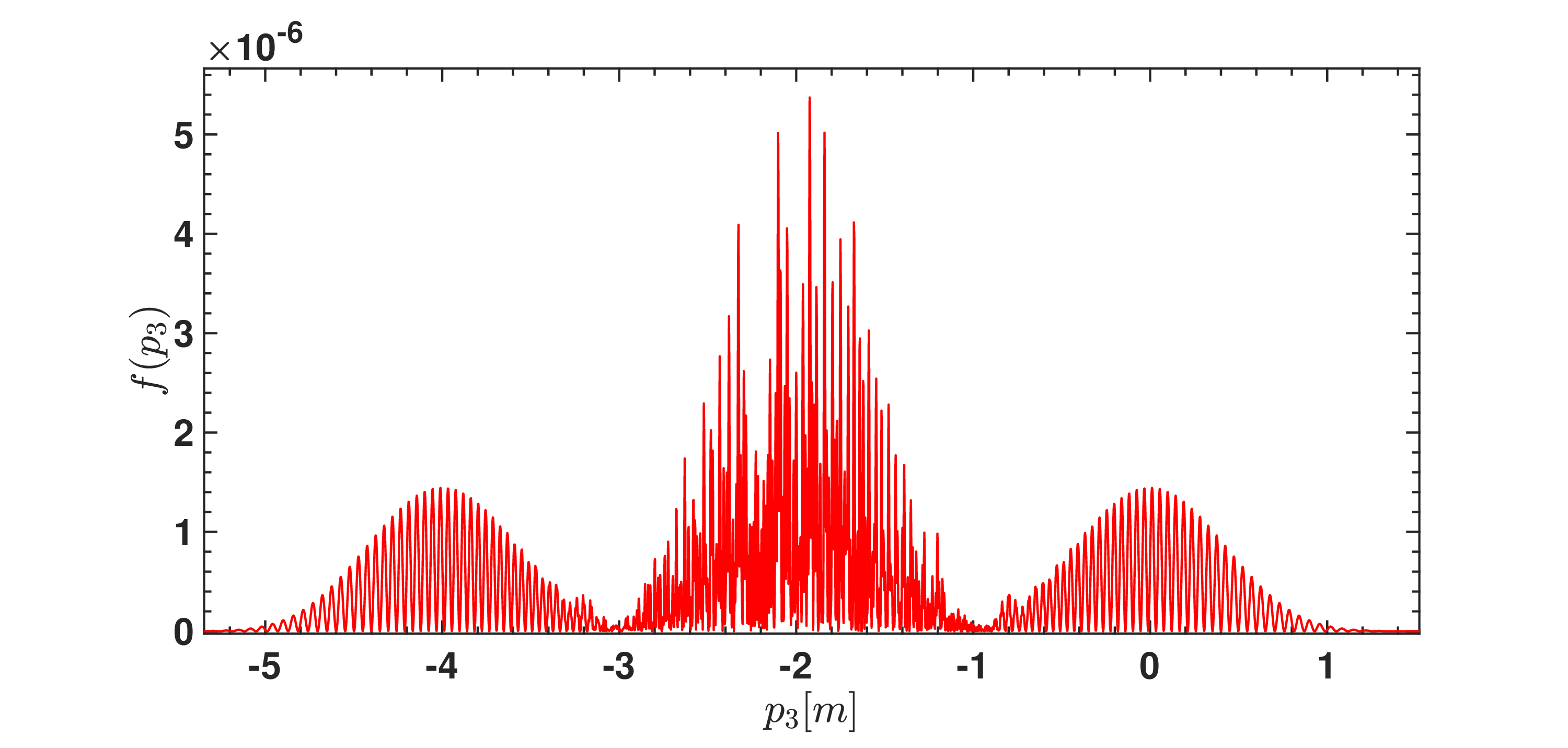}}
\caption{\label{Fig6} Same as Fig.~\ref{Fig1} but for $\xi_{12} =2.00$}
\end{figure}

%%%

For larger asymmetry,$\xi_{12} =1.5$ and $2.0$ (Figs.~\ref{Fig5} and \ref{Fig6}), the spectra undergo a qualitative transition. 
At $\xi_{12} =1.5$, the periodic sequence develops a multi-peak oscillatory structure with enhanced amplitude 
$ f(p_3)\sim 10^{-9}$, indicating partial constructive interference across multiple momentum channels. The quasiperiodic spectrum, while still oscillatory, shows fewer dominant peaks with fine substructure and a reduced coherence buildup. 

At $\xi_{12} =2.0$, the periodic spectrum collapses into a small number of sharply localized resonant peaks, 
indicating strong momentum selectivity. In contrast, the quasiperiodic sequence exhibits a broader distribution with 
enhanced amplitudes ($ \sim 10^{-6}$) and reduced sensitivity to precise phase matching.
Overall, these results demonstrate that long-range temporal order plays a decisive role in controlling both spectral 
structure and production efficiency. Periodic driving supports coherent resonances and strong momentum selectivity,
while quasiperiodic ordering redistributes spectral weight  and can mitigate destructive interference, depending on
$\xi_{12}$.

\begin{figure}[tbp]
\centering
{\includegraphics[width=0.9682\columnwidth]{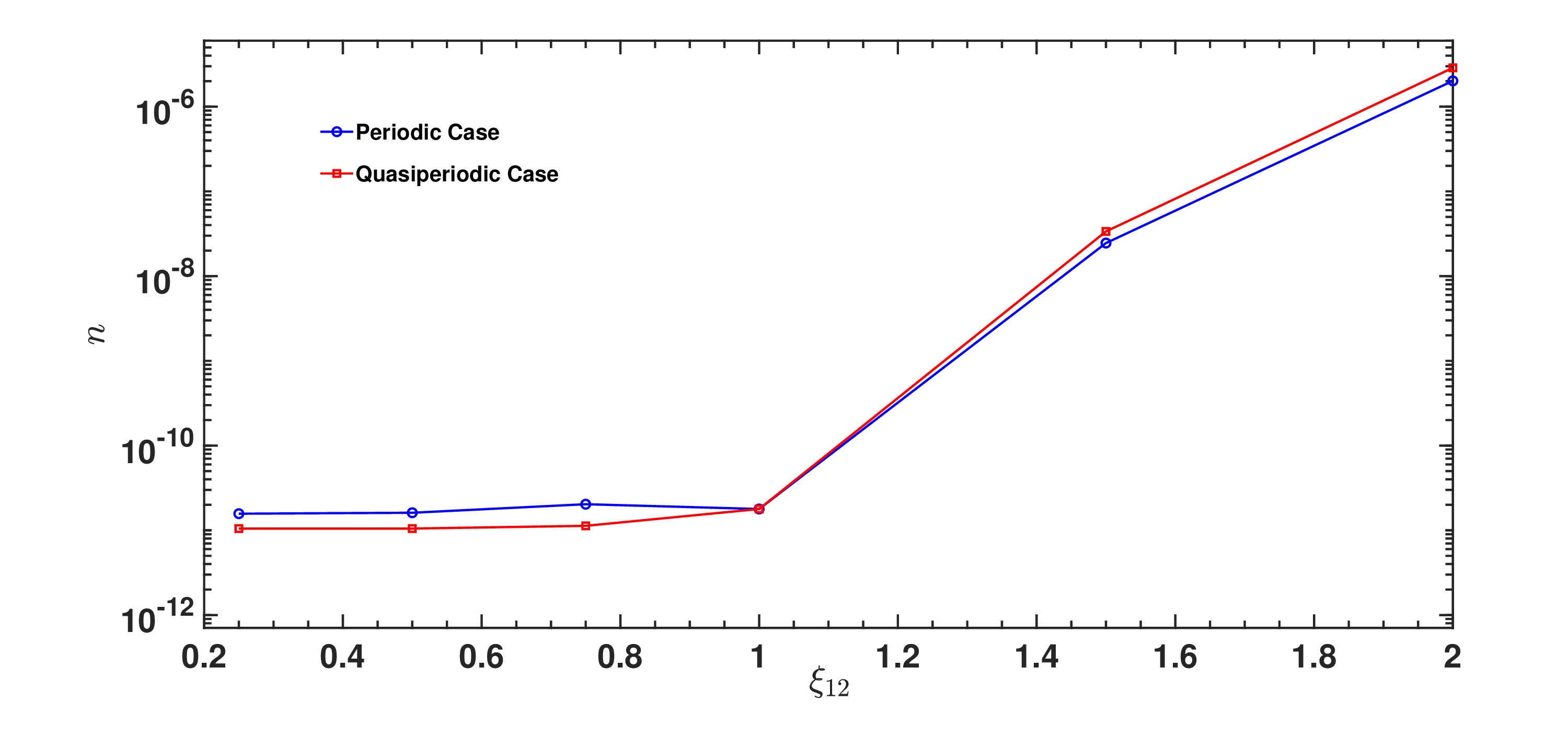}}
\caption{\label{Fig7} 
Particle yield $n$ as a function of the 
field-strength ratio $\xi_{12}$ for $N=12$ pulses. 
Blue curve: periodic sequence; red curve: Fibonacci quasiperiodic sequence.
}
\end{figure}

\par

The momentum spectra discussed in Figs.~\ref{Fig1}–\ref{Fig6} reveal rich interference structures whose complexity depends sensitively on both $\xi_{12}$ and the temporal ordering. 

%To quantify the overall efficiency of each driving scheme, we now examine the integrated particle yield $n(+\infty)$, which sums contributions from all longitudinal momenta.

Finally, Fig.~\ref{Fig7} shows the integrated particle  yield $n(+\infty)$ as a function of the field-strength ratio \(\xi_{12}  = E_1/E_2\) for \(N = 12\) pulses. 
In contrast to the rich interference structures observed in the momentum spectra, the total yields exhibit a comparatively smooth and monotonic dependence on $\xi_{12}.$

%%%
For both periodic and quasiperiodic sequences, the yield remains strongly suppressed in the regime $\xi_{12} <1,$ with values of order $10^{-11} -10^{-10}.$  In this region, the two curves are nearly indistinguishable, indicating that temporal ordering has only a minor influence on the overall production efficiency when the field amplitudes are weak or comparable.
A sharp increase in the particle yield is observed as $\xi_{12}$ exceeds unity, signaling the onset of efficient pair production driven by the stronger field component. In this regime, both sequences exhibit a similar  scaling behavior, with the yield  rising by several orders of magnitude as $\xi_{12}$ increases from $1$ to $2.$
%%%
A small but systematic difference emerges between the two cases: for $\xi_{12} >1 $, the quasiperiodic (Fibonacci) sequence produces a slightly higher yield than the periodic sequence, whereas for $\xi_{12} <1,$ the periodic case is marginally larger. However, this difference remains modest across the entire
parameter range, indicating that temporal ordering plays a secondary role in determining the yield compared to the overall field strength.
These results demonstrate that, although periodic and quasiperiodic sequences produce qualitatively distinct momentum spectra, their particle yields are remarkably similar. This highlights that the total yield is governed primarily by the effective field strength, while the temporal ordering predominantly influences the redistribution of particles in momentum space rather than the overall production rate.

\subsection{Effects of increasing pulse number: N=20}

When the number of pulses is increased from \(N = 12\) to \(N = 20\),  while keeping all other parameters fixed (\(E_2 = 0.1 E_{c}\), \(T = 78.2 [m^{-1}]\)), the interference structures in the momentum spectra becomes sharper and more pronounced. This reflects the increased accumulation of dynamical phase over a longer pulse train, leading to improved resolution of interference features, analogous to increasing the number of slits in a multi-slit interferometer.

%can be understood as analogous to using a finer ruler or a larger antenna array: the resolution improves, and the interference effects become more extreme.

We now examine how increasing the pulse number to $N = 20$ modifies the longitudinal momentum spectra and amplifies the role of temporal ordering. Throughout this subsection, periodic and Fibonacci pulse sequences are compared under identical field parameters.

\begin{figure}[tbp]
\centering
{\includegraphics[width=0.9682\columnwidth]{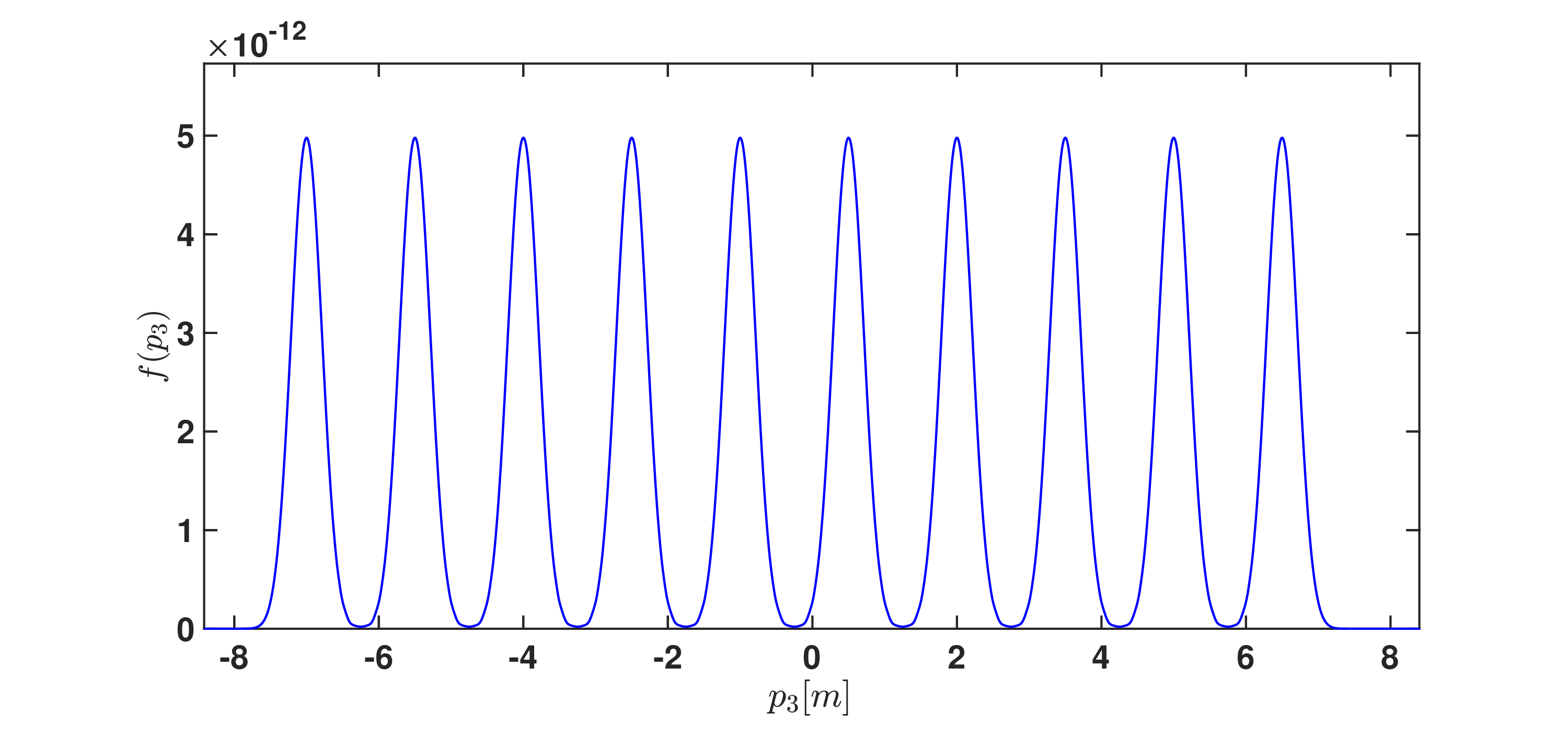}
\includegraphics[width=0.9682\columnwidth]{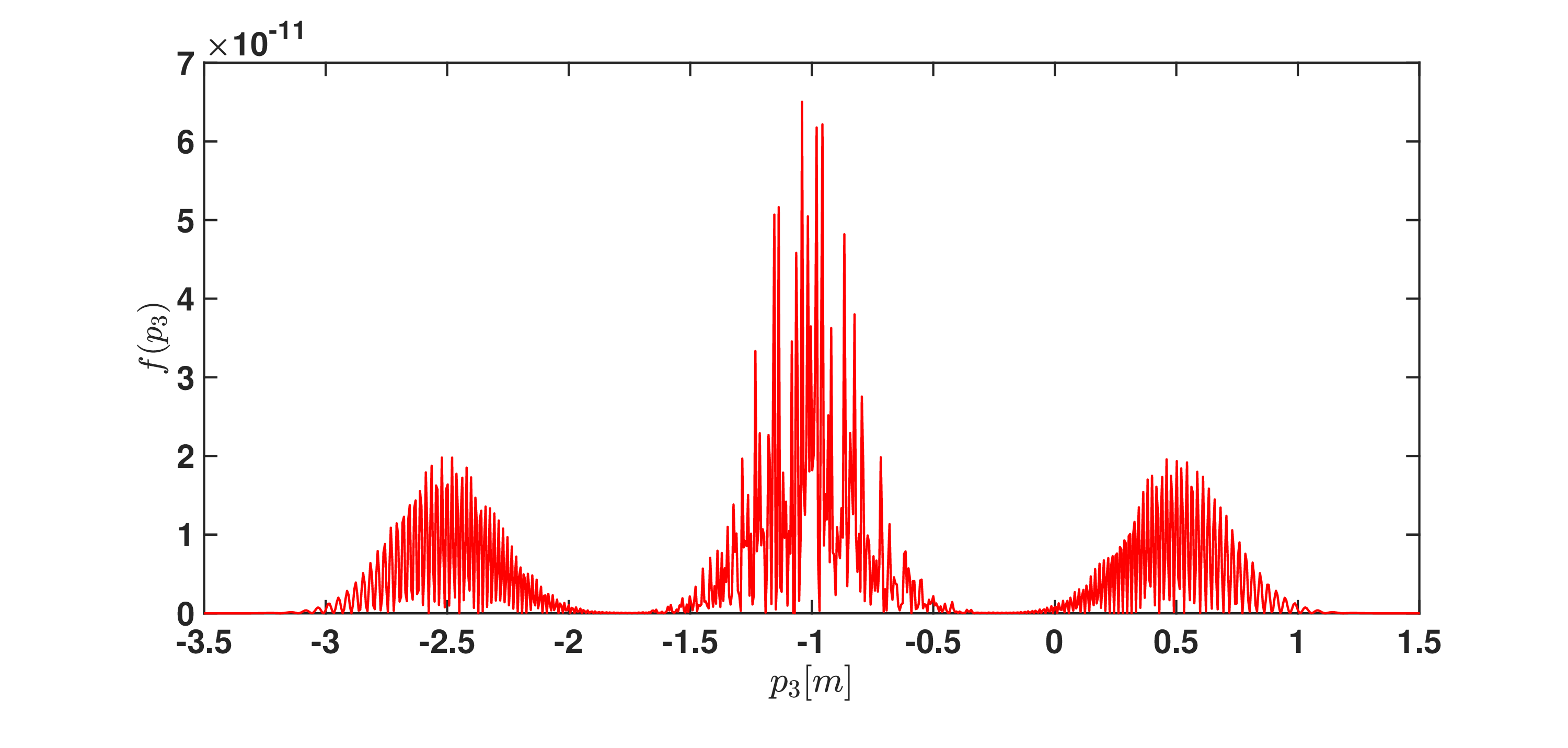}}
\caption{\label{Fig8} Longitudinal momentum spectrum for $N=20$  at $\xi_{12}=0.25$
Upper panel : Periodic sequence ; Lower Panel : Fibonacci sequence.}
\end{figure}

Figure~\ref{Fig8} shows the longitudinal momentum spectrum for \(N = 20\) at \(\xi_{12} = 0.25\). Compared to the  \(N = 12\) case, increasing the pulse number primarily sharpens interference-induced structures without 
significantly increasing the overall amplitude.

%%%%
%%%
For the periodic sequence[upper panel of Fig.~\ref{Fig8}], the spectrum develops a more 
regular oscillatory pattern extending over a wider momentum range.
 The increased pulse number enhances temporal coherence, leading to narrower and better-resolved interference fringes. 
 However, the overall magnitude remains strongly suppressed due to the imbalance between the two field amplitudes.

In the Fibonacci sequence, the spectrum becomes more localized in momentum space, with reduced sidebands and diminished support at larger $|p_3|.$ This indicates that quasiperiodicity is increasingly  inhibits long-range phase coherence as the pulse train length grows.

%The Fibonacci sequence exhibits enhanced momentum-space confinement relative to \(N = 12\). The spectrum becomes more narrowly localized, with suppressed sidebands and reduced spectral support at large \(|p_3|\). This behavior indicates that quasiperiodicity increasingly inhibits long-range phase coherence as the pulse train length grows, even in the weak-modulation regime.

\begin{figure}[tbp]
\centering
{\includegraphics[width=0.9682\columnwidth]{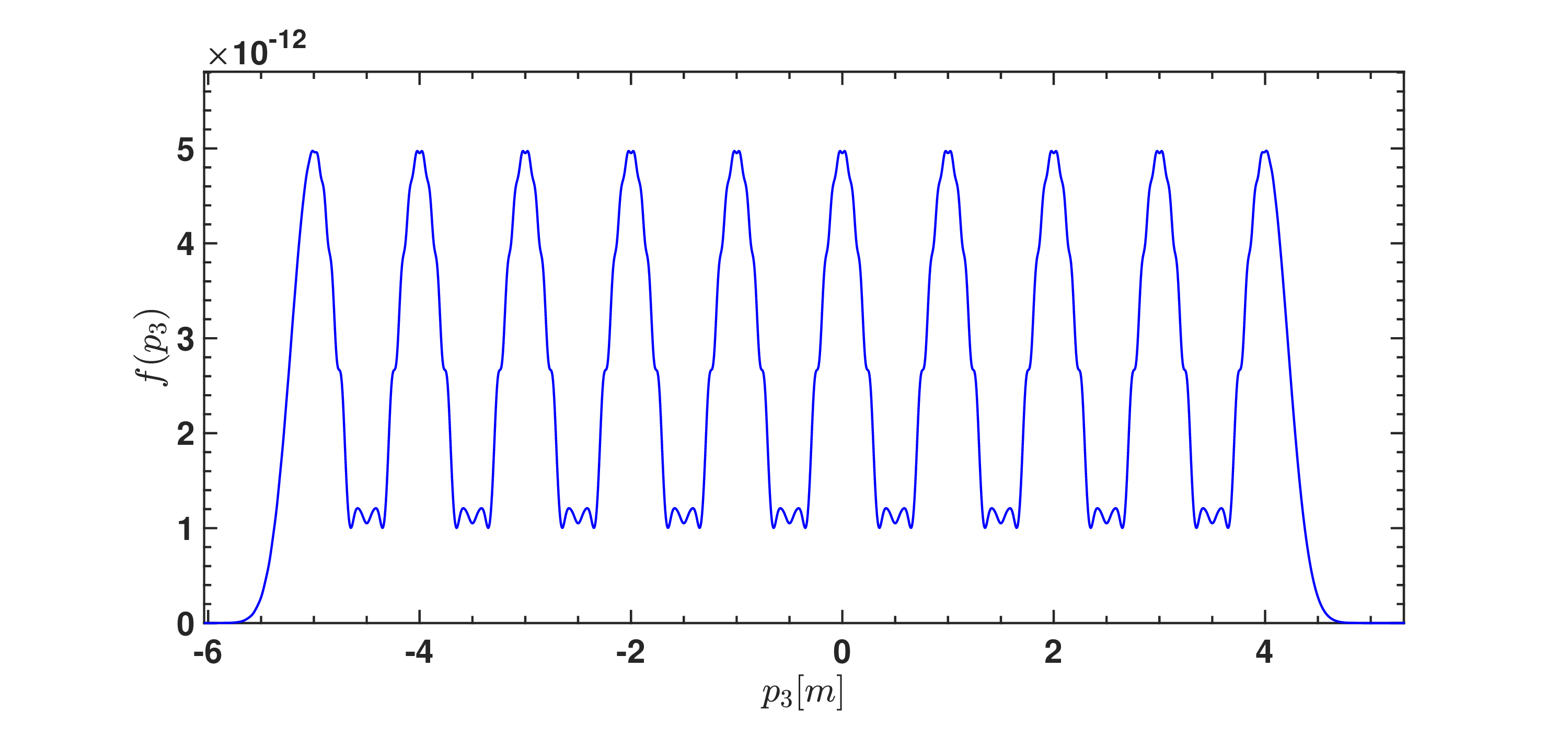}
\includegraphics[width=0.9682\columnwidth]{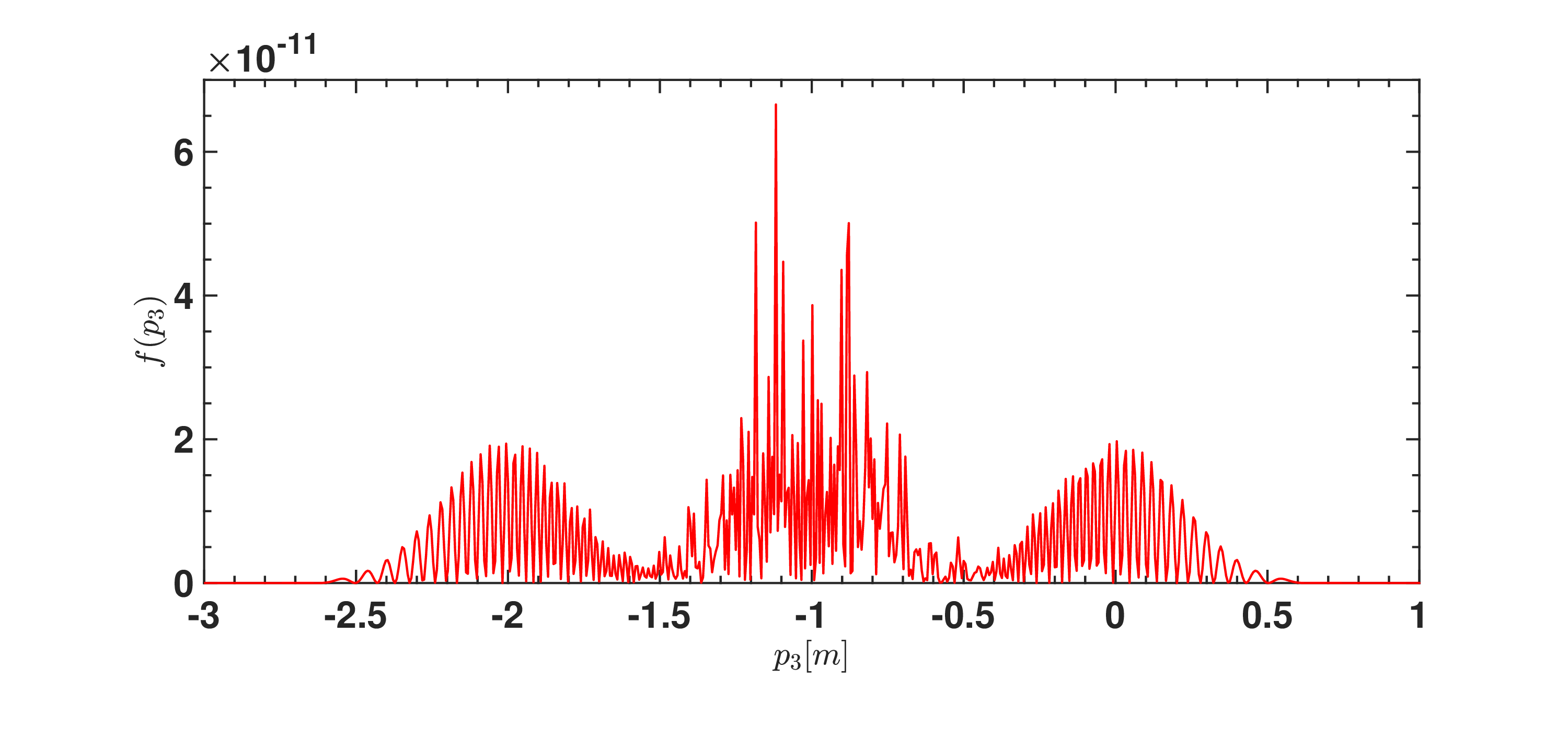}
}
\caption{\label{Fig9} Same as Fig.~\ref{Fig8} but for \(\xi_{12} = 0.50\)}
\end{figure}

At \(\xi_{12} = 0.50\) (Fig.~\ref{Fig9}), the influence of temporal ordering becomes more pronounced. The periodic spectrum exhibits sharper and higher-contrast peaks, reflecting stronger constructive interference across the pulse train.
In contrast, the Fibonacci spectrum shows further localization, with spectral weight concentrated in a narrower momentum 
interval and stronger suppression outside this region. This reflects the cumulative phase mismatch introduced by
the quasiperiodic structure.

%In this regime, the effects of temporal ordering become substantially more pronounced than for \(N = 12\).

%The periodic case spectrum retains its extended support and regular modulation, but with significantly sharper and higher-contrast peaks. The enhanced pulse number strengthens constructive interference across the pulse train, resulting in a clearer and more refined interference comb (see Fig.~\ref{Fig9}).

%In contrast, the \textit{Fibonacci} spectrum undergoes further localization. Spectral weight is concentrated into a narrower momentum interval, with sharper peaks and stronger suppression outside this region. The redistribution of spectral weight observed at \(N = 12\) evolves into a clear momentum-space confinement at \(N = 20\), reflecting the cumulative phase mismatch generated by the quasiperiodic temporal structure.

\begin{figure}[tbp]
\centering
{\includegraphics[width=0.9682\columnwidth]{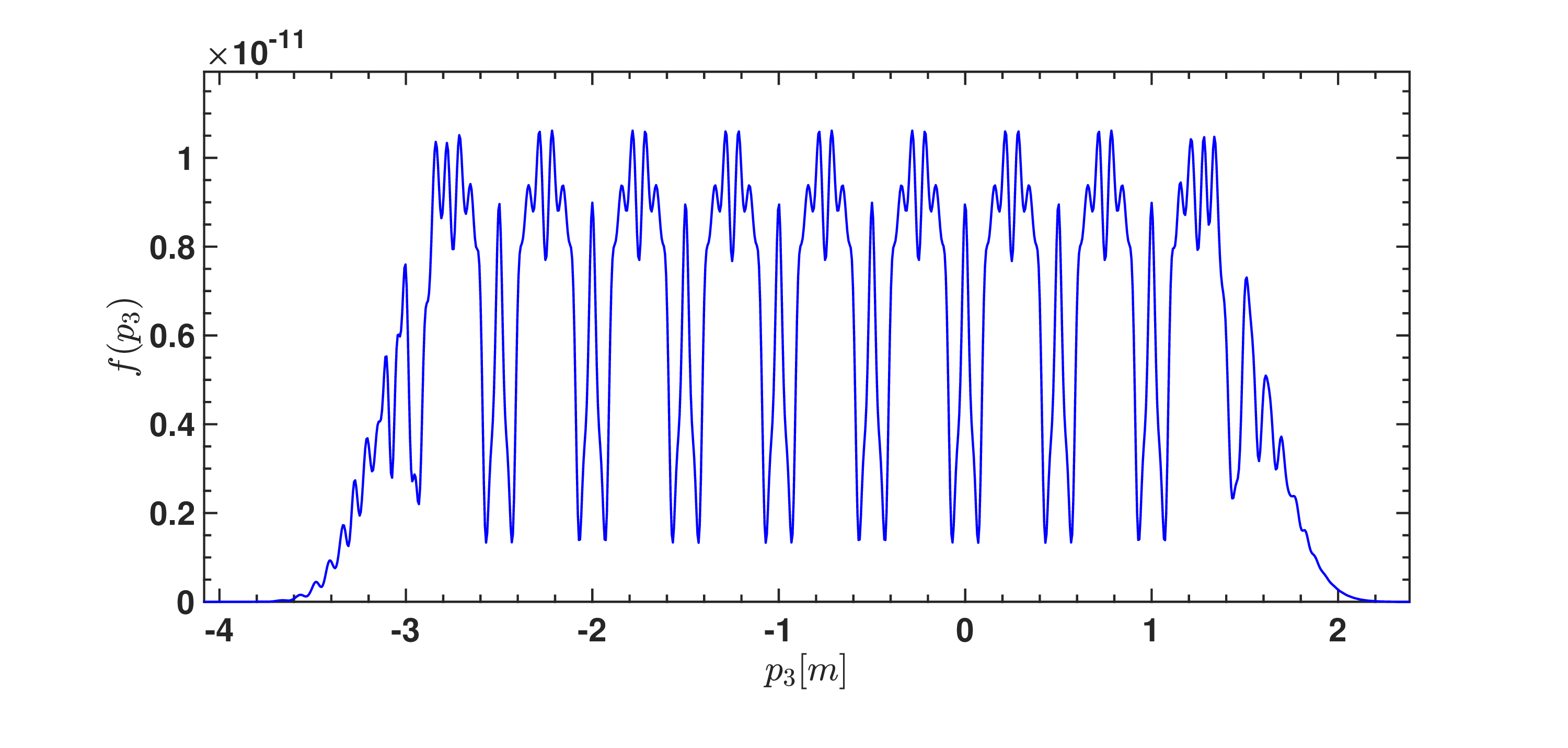}
\includegraphics[width=0.9682\columnwidth]{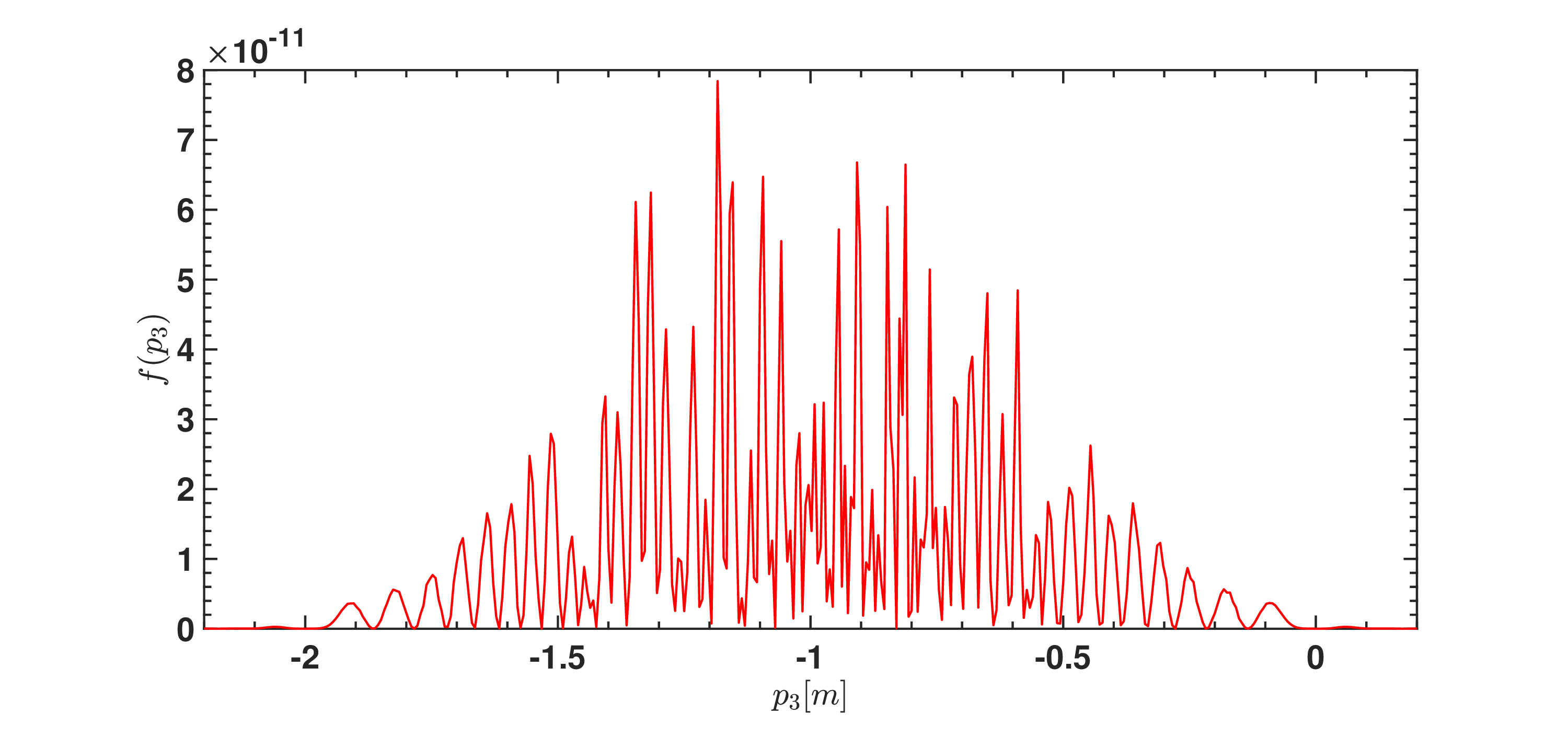}
}
\caption{\label{Fig10} Same as Fig.~\ref{Fig8} but for \(\xi_{12} = 0.75\)}
\end{figure}

Figure~\ref{Fig10} shows the spectra for \(\xi_{12} = 0.75\).
The periodic sequence maintains a broad distribution with regular interference fringes, indicating sustained phase coherence despite overall suppression.

%a regime that already exhibited strong fragmentation for Fibonacci ordering at \(N = 12\).

%For the periodic sequence, the spectrum remains broadly distributed with well-defined oscillatory fringes, consistent with coherent multipulse interference. Although the overall yield is suppressed, the regular structure indicates preserved phase coherence.

The Fibonacci spectrum, however, becomes increasingly fragmented, forming a dense set of irregular peaks confined to a narrow momentum window.
Compared to $N=12,$ these structures are sharper and more numerous, indicating enhanced sensitivity to quasiperiodic temporal ordering.

%collapses into a dense cluster of irregular, high-contrast peaks confined to a narrow momentum window. Compared to \(N = 12\), the peaks are sharper and more numerous, demonstrating enhanced sensitivity to incommensurate temporal scales as the pulse number increases. This behavior signals the strengthening of quasiperiodicity-induced phase decoherence rather than its averaging out.

\begin{figure}[tbp]
\centering
{\includegraphics[width=0.9682\columnwidth]{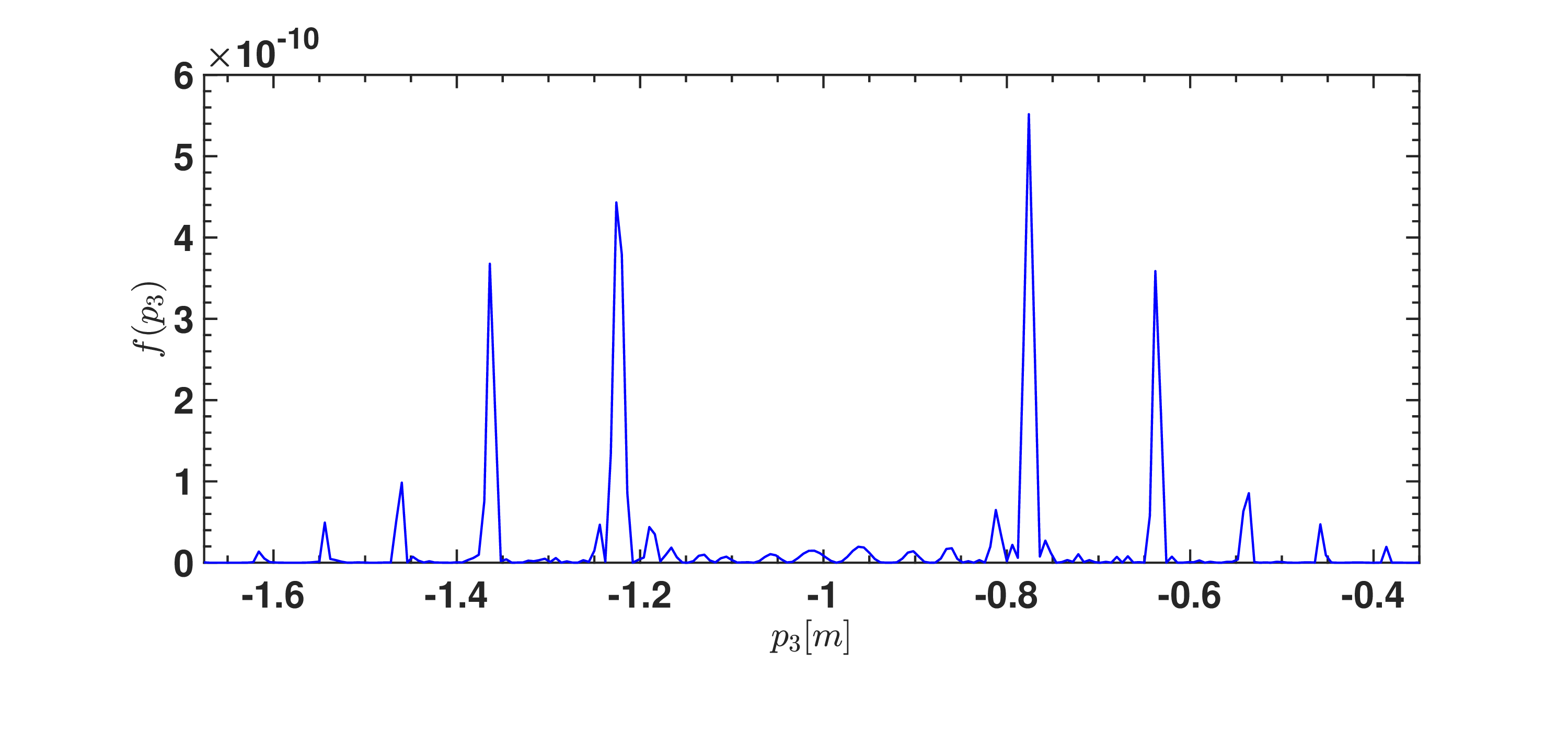}
\includegraphics[width=0.9682\columnwidth]{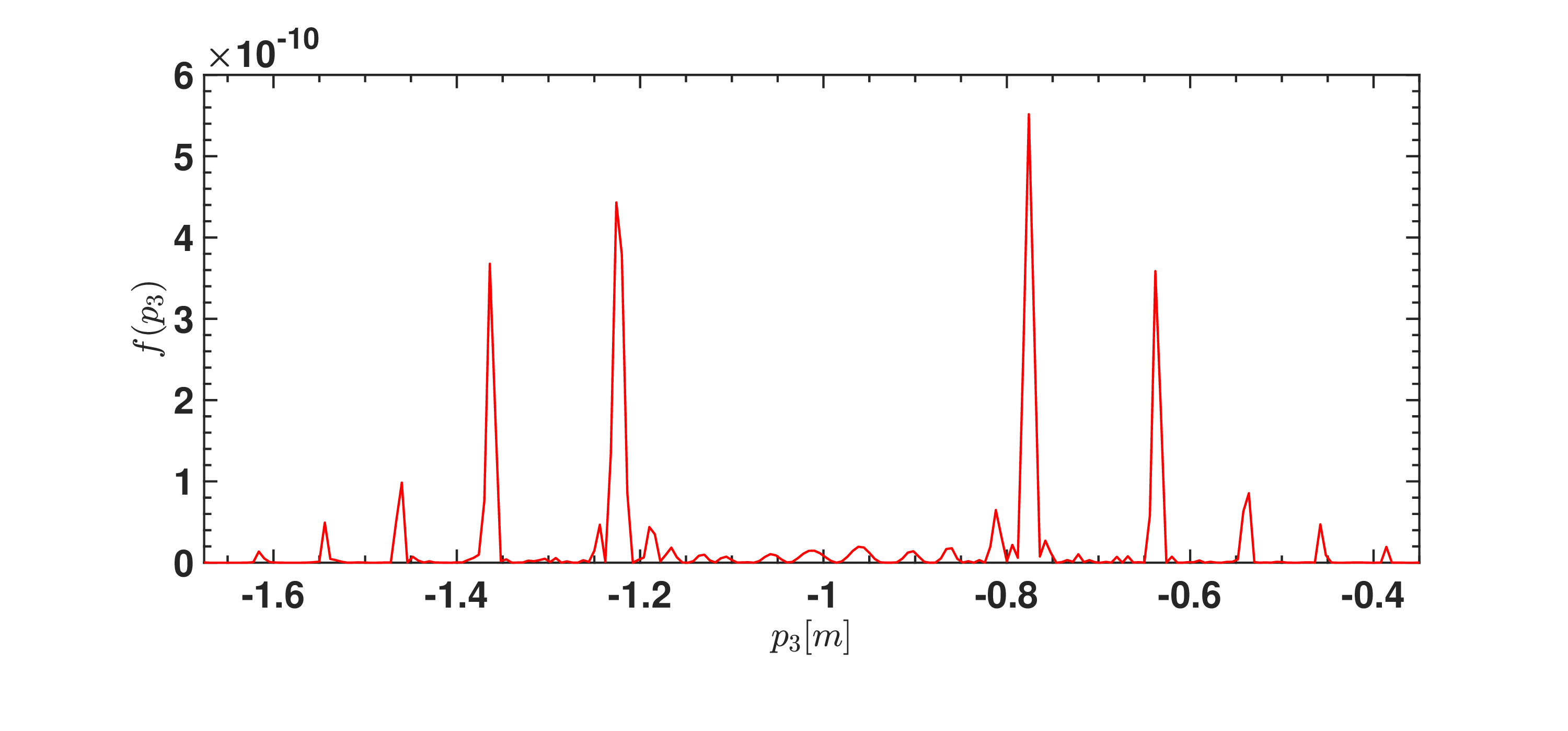}
}
\caption{\label{Fig11} Same as Fig.~\ref{Fig8} but for \(\xi_{12} = 1.0\)}
\end{figure}

%%The case

At \(\xi_{12} = 1.0\)  (Fig.~\ref{Fig11}), corresponding to equal pulse amplitudes,  the periodic spectrum exhibits a highly regular and sharply resolved interference pattern over a wide momentum range.
The peaks are narrow and evenly spaced, indicating strong global phase coherence. Increasing $N$ further enhances peak sharpness and contrast.
\par
In contrast, the Fibonacci sequence produces a more localized spectrum dominated by a few prominent peaks, with strong 
suppression elsewhere. This indicates that quasiperiodicity restricts constructive interference to limited momentum regions.

%For the periodic pulse train[Fig.~\ref{Fig11}], the momentum spectrum exhibits a highly regular and sharply resolved interference pattern extending over a broad momentum range. The peaks are narrow, evenly spaced, and of comparable height, indicating robust global phase coherence across the entire \(N = 20\) pulse sequence. Relative to \(N = 12\), the increased pulse number leads to further peak narrowing and enhanced contrast, consistent with strengthened temporal coherence.

%In stark contrast, the Fibonacci sequence produces a strongly localized spectrum dominated by a few intense, isolated peaks, with near-complete suppression elsewhere. Compared to the \(N = 12\) case, localization effects are significantly amplified: the peaks become sharper, the background is further reduced, and momentum selectivity is enhanced. This confirms that quasiperiodicity-induced phase decoherence accumulates with pulse number, restricting constructive interference to a small number of resonant momenta.

\begin{figure}[tbp]
\centering
{\includegraphics[width=0.9682\columnwidth]{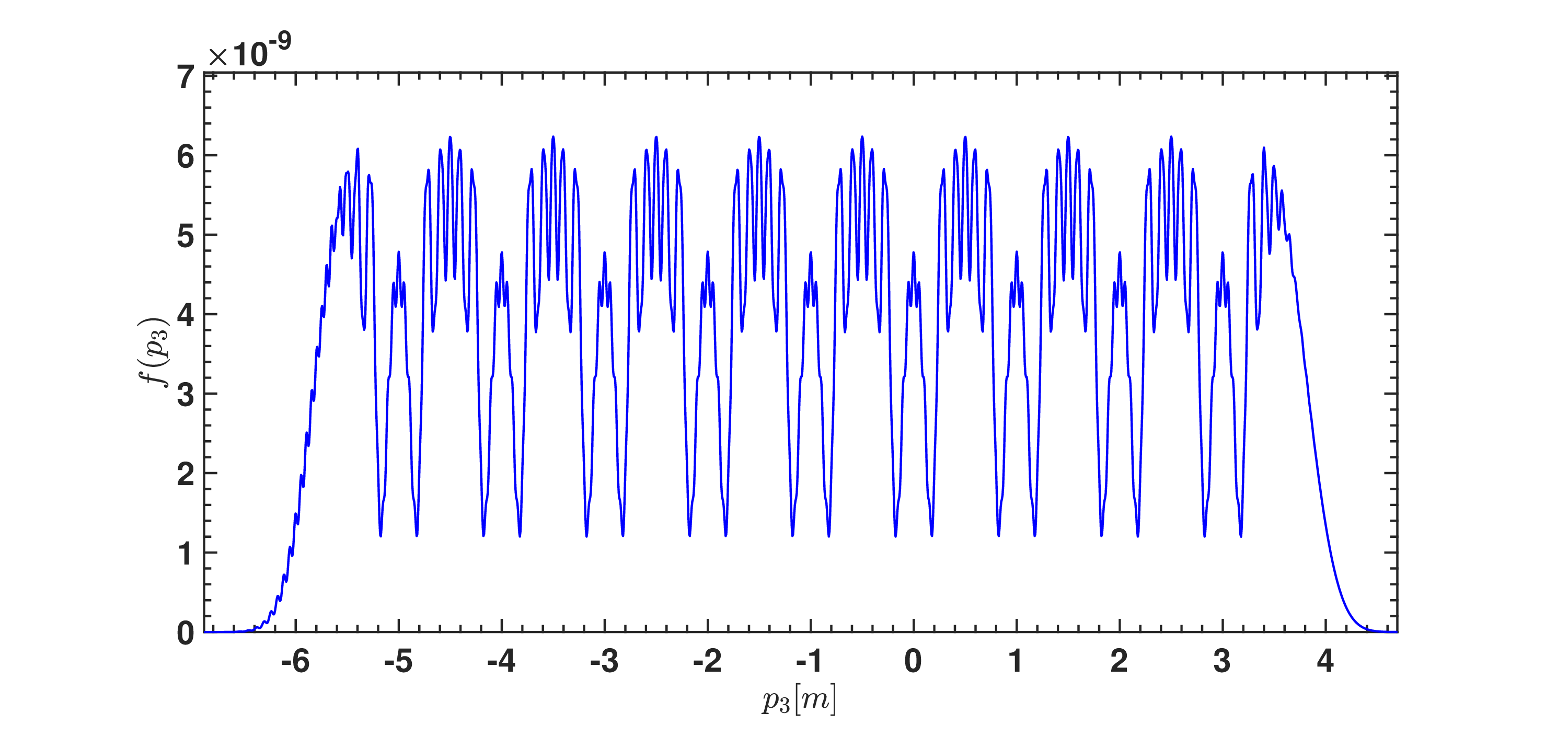}
\includegraphics[width=0.9682\columnwidth]{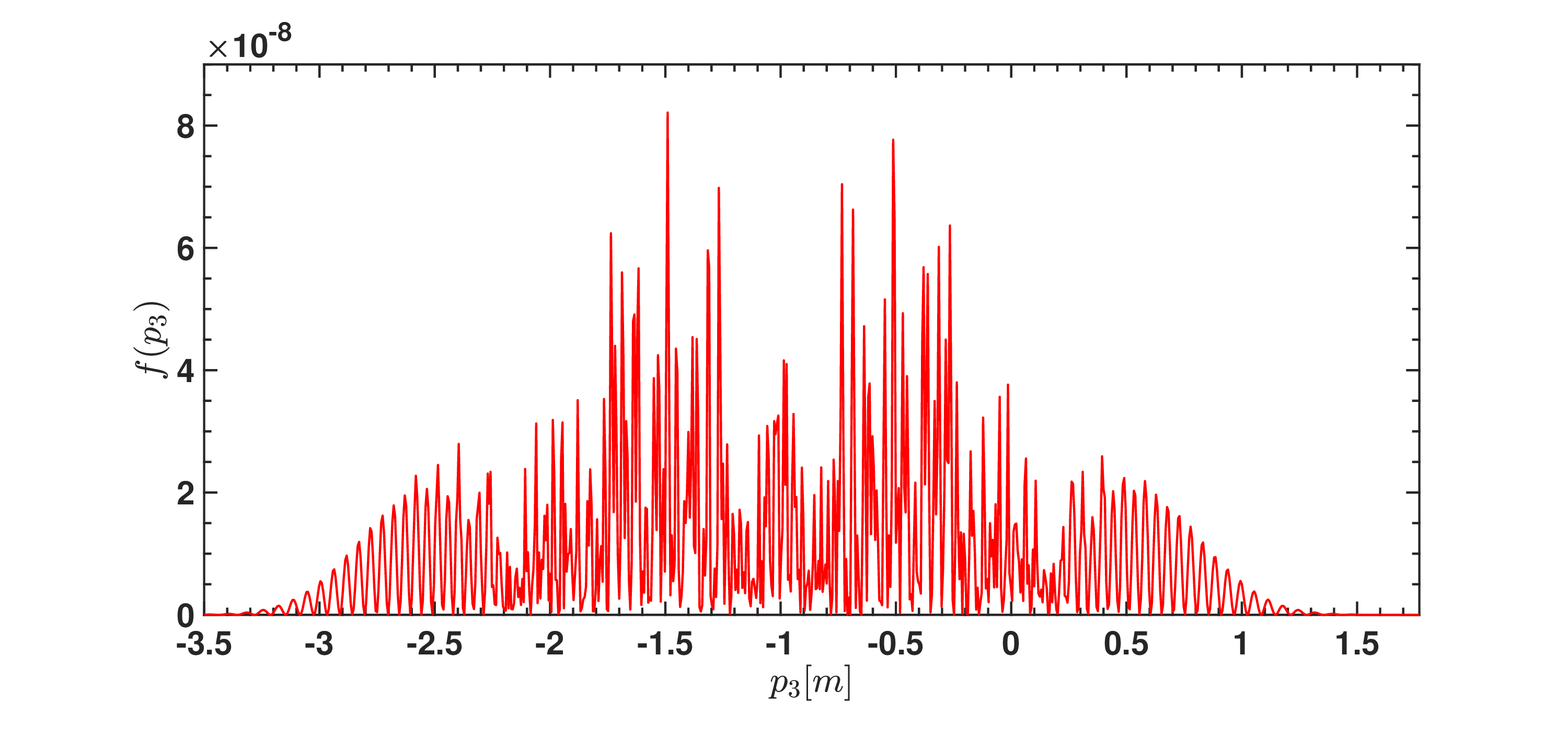}}
\caption{\label{Fig12} Same as Fig.~\ref{Fig8} but for \(\xi_{12} = 1.5\)}
\end{figure}

%Figure~\ref{Fig12} shows the spectra at \(\xi_{12} = 1.5\). At this larger field-strength ratio, both sequences exhibit increased overall yield, but their spectral characteristics remain markedly different.

At $\xi_{12} =1.5$ (Fig.~\ref{Fig12}), both sequences exhibit increased overall amplitude. The periodic spectrum retains its regular comb-like structure, reflecting sustained constructive interference.

The Fibonacci spectrum shows irregular peak clustering with varying amplitudes and spacing, indicating the presence of 
multiple competing interference contributions arising from quasiperiodic ordering.

%The periodic spectrum increases in magnitude while preserving its smooth, comb-like interference structure, reflecting constructive interference maintained over long temporal scales (see Fig.~\ref{Fig12}).

%The Fibonacci spectrum, however, displays dense peak clustering with strong fluctuations in peak height and irregular spacing. Compared to \(N = 12\), localization is stronger and the spectral envelope is more sharply defined, indicating the presence of multiple competing resonant pathways generated by quasiperiodic ordering.

\begin{figure}[tbp]
\centering
{\includegraphics[width=0.9682\columnwidth]{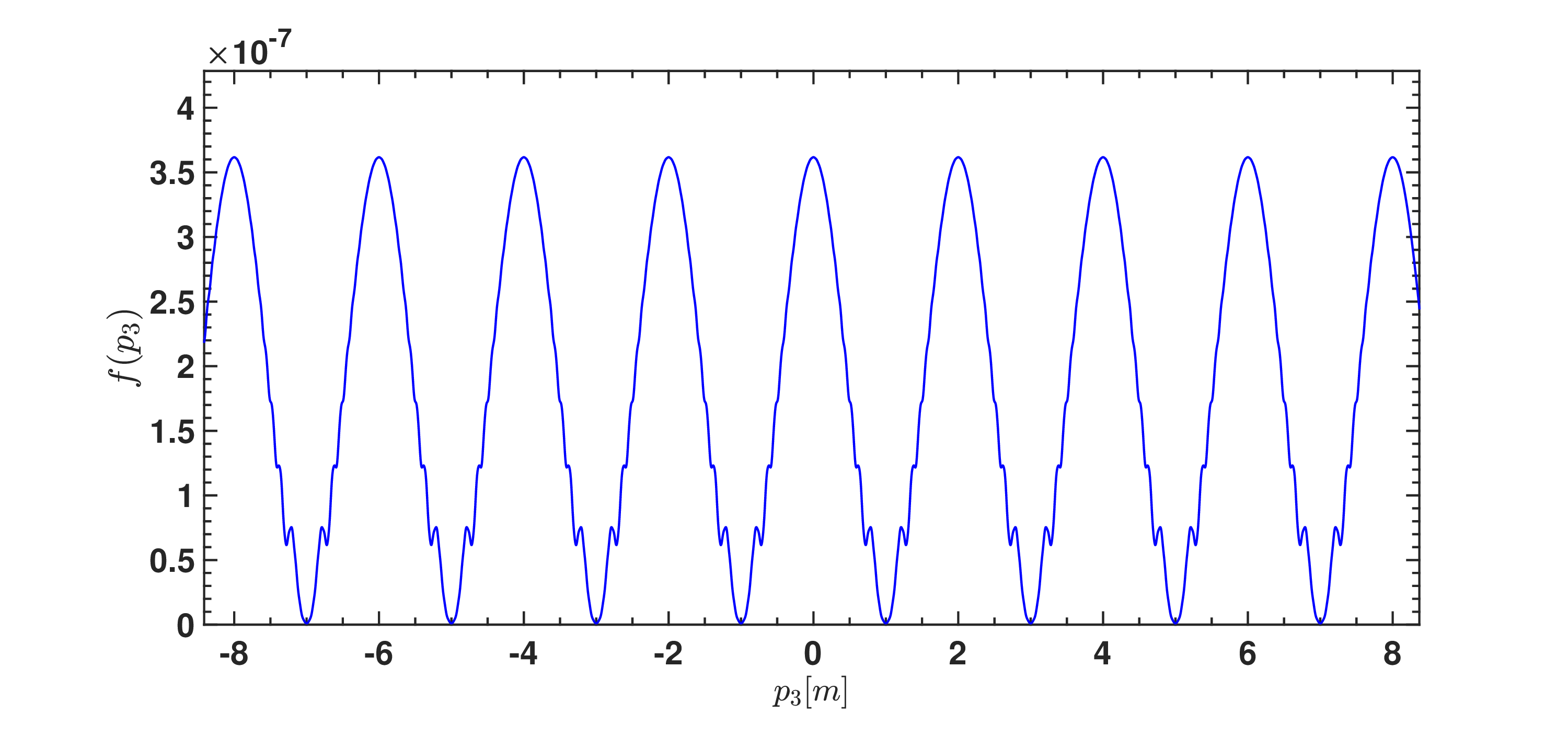}
\includegraphics[width=0.9682\columnwidth]{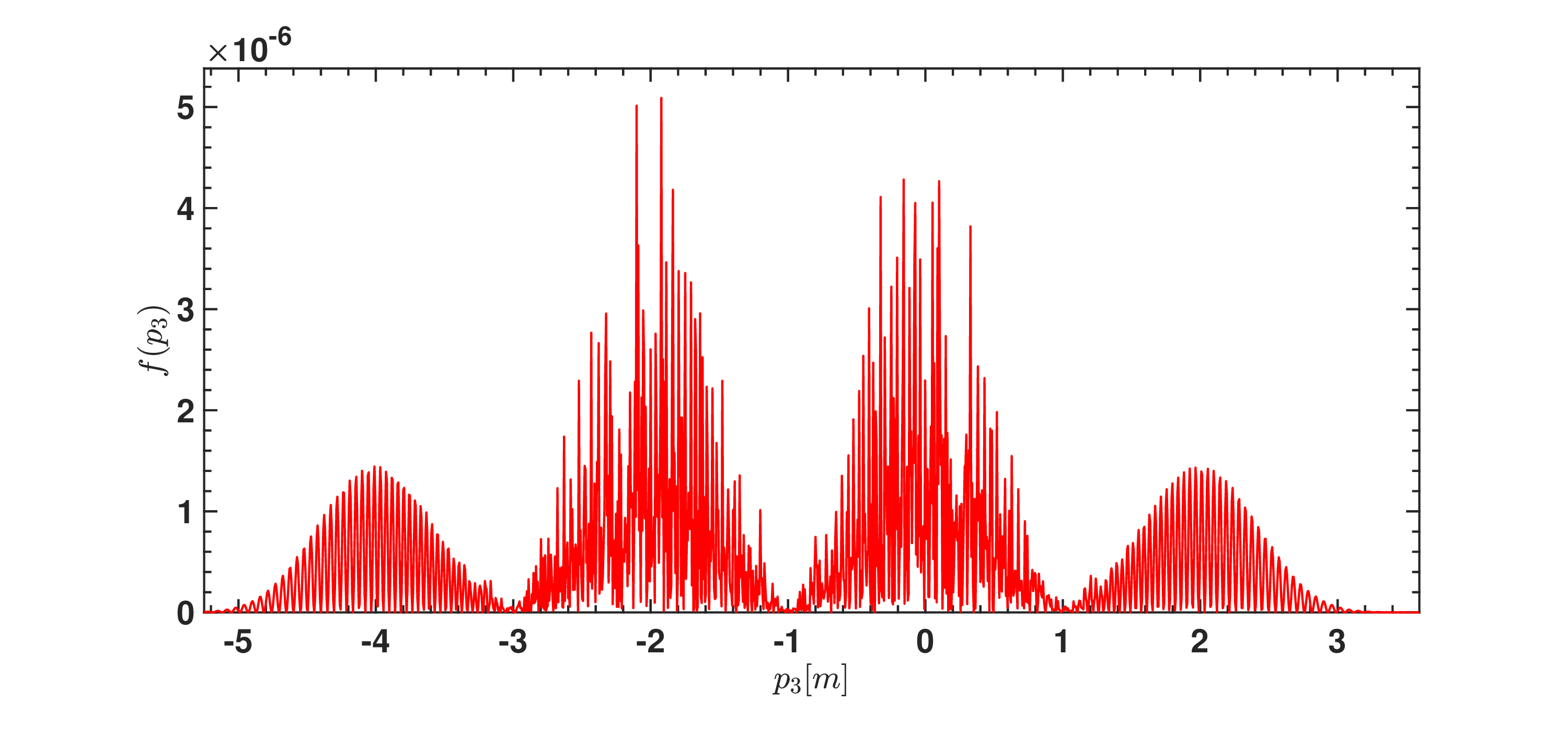}
}
\caption{\label{Fig13} Same as Fig.~\ref{Fig8} but for \(\xi_{12} = 2.0\)}
\end{figure}

%At the largest field-strength ratio, shown in Fig.~\ref{Fig13}, the cumulative effects of temporal ordering become most evident.

At $\xi_{12} =2.0$ (Fig.~\ref{Fig13}), the periodic spectrum remains extended and regularly modulated, indicating stable phase coherence even for large $N.$

In contrast, the Fibonacci spectrum becomes more strongly localized, with sharp peaks confined to specific momentum regions and suppression elsewhere. This demonstrates that increasing the pulse number enhances quasiperiodicity -induced momentum filtering.
%%%
Overall, increasing the pulse number from $N=12$ to $N =20$ does not change the qualitative distinction between periodic
and quasiperiodic sequences, but it amplifies their differences. Periodic driving leads to increasingly 
sharp and coherent interference patterns, while quasiperiodic ordering promotes localization and spectral 
fragmentation. The dependence on $\xi_{12}$ becomes more pronounced at larger $N,$ indicating that longer pulse trains 
provide enhanced control over the momentum distribution of produced pairs through accumulated phase effects.

\begin{figure}[tbp]
\centering
{\includegraphics[width=0.9682\columnwidth]{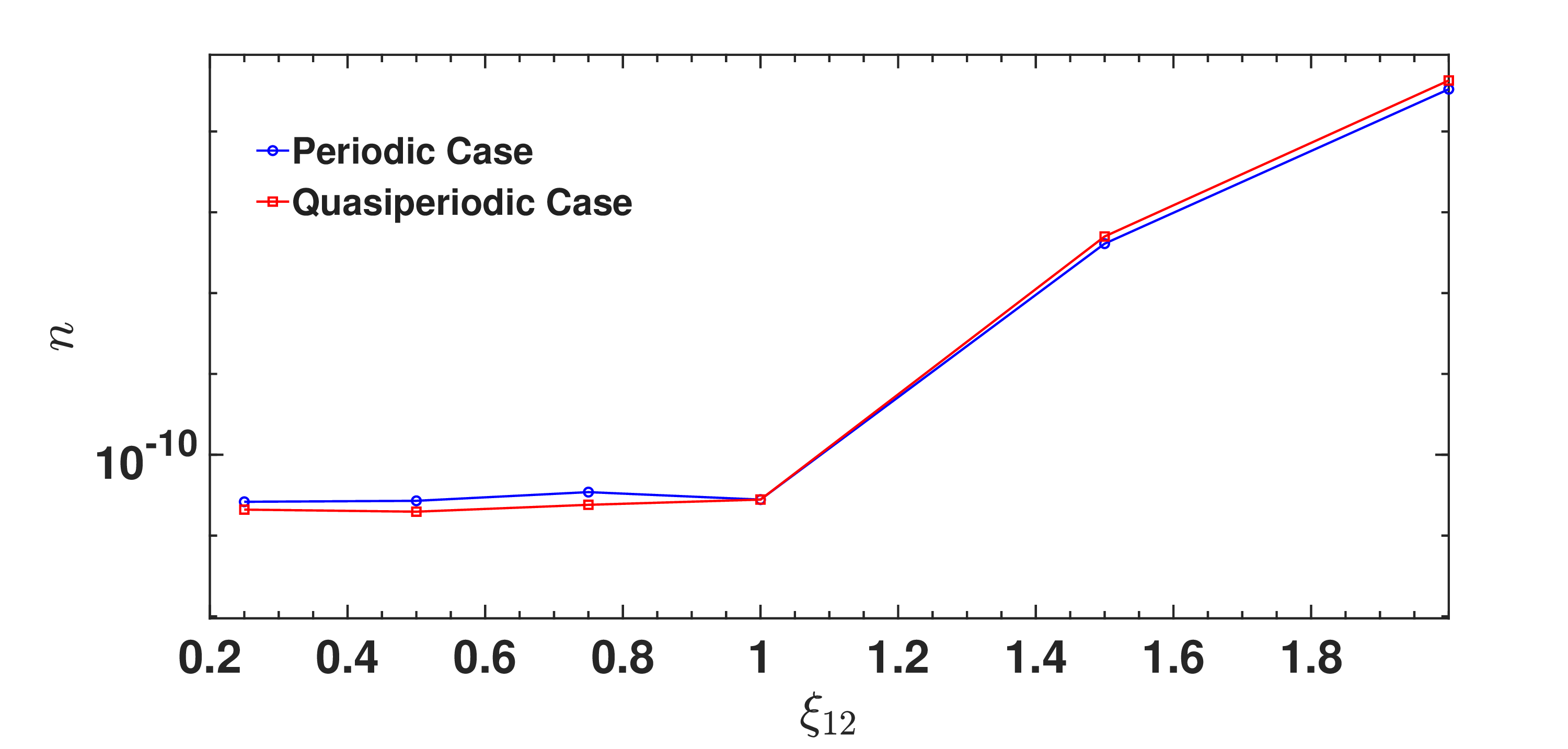}}
\caption{\label{Fig15} Particle yield $\mathcal{n}(+\infty)$ as a function of $\xi_{12}$ for $N=20$ pulses. Blue curve:periodic sequence; red curve : Fibonacci quasiperiodic sequence}
\end{figure}

Figure~\ref{Fig15} shows the total particle yield $\mathcal{n}(+\infty)$ as a function of the field-strength ratio $\xi_{12}$ for $N = 20$ pulses. This provides an integrated measure of pair-production efficiency and complements the momentum-resolved spectra presented in Figs.~\ref{Fig8}--\ref{Fig13}.

As in the $N = 12$ case, the particle yield exhibits a smooth and strongly nonlinear dependence on $\xi_{12}$. For $\xi_{12} < 1$, the yield remains strongly suppressed for both periodic and quasiperiodic sequences, with only minor differences between the two. This indicates that, in the weak-modulation regime, temporal ordering has a limited impact on the overall production efficiency.

As $\xi_{12}$ increases beyond unity, the particle yield rises rapidly by several orders of magnitude for both sequences, reflecting the increasing dominance of the stronger field component. Increasing the number of pulses from $N = 12$ to $N = 20$ leads to an overall enhancement of the yield, consistent with cumulative multipulse effects and coherent phase accumulation.

Despite the pronounced differences observed in the momentum spectra, the integrated yields for periodic and quasiperiodic sequences remain relatively close across the entire parameter range. A modest but systematic difference is observed: the quasiperiodic sequence tends to produce slightly higher yields for $\xi_{12} > 1$, while the periodic sequence is marginally favored for $\xi_{12} < 1$. However, these differences remain within the same order of magnitude.

Importantly, no sharply resolved resonant structures are observed in the integrated yield. This indicates that the strong interference effects visible in momentum space do not directly translate into equally strong variations in the total particle number. Instead, temporal ordering primarily redistributes spectral weight in momentum space, while the total yield is governed predominantly by the effective field strength.

These results demonstrate that increasing the pulse number amplifies the overall production efficiency but does not qualitatively change the role of temporal ordering in determining the integrated yield. Periodic and quasiperiodic sequences therefore, exhibit a similar global scaling behavior, despite their markedly different momentum-space signatures.

\section{Summary and outlook}

%%%%%20march%%%

In this work, we investigate nonperturbative electron–positron pair production from the quantum vacuum driven by alternating-sign multi-pulse electric field trains with periodic and quasiperiodic amplitude modulation. The influence of the pulse number and the arrangement of field strengths is studied by solving the quantum Vlasov equation.

We show that long-range temporal order in the driving field serves as a key control parameter. It strongly influences the momentum-resolved spectra of produced particles, generating distinct interference patterns, while its impact on the particle yield remains comparatively moderate. This demonstrates that temporal pulse engineering primarily governs spectral redistribution in momentum space, whereas the overall production rate is dominated by the effective field strength.

A central result of our analysis is that the longitudinal momentum distribution \(f(p_3)\) acts as a sensitive interferometric probe of the underlying pulse sequence. For periodic \((E_1, E_2)\) pulse trains, the spectrum exhibits high-contrast interference patterns reminiscent of a multi-slit-in-time configuration, with peak sharpness and contrast increasing with the number of pulses \(N\). This behavior reflects coherent quantum interference across the pulse train. In contrast, Fibonacci quasiperiodic sequences generate fragmented and irregular spectral structures, indicating the absence of global phase coherence. This fragmentation becomes more pronounced as the pulse number increases, demonstrating that quasiperiodic ordering redistributes spectral weight in momentum space rather than concentrating it into well-defined interference peaks.

We further identified the field-strength ratio \(\xi_{12} = E_1/E_2\) as a key control parameter that determines the interference regime. For periodic driving, variations in \(\xi_{12}\) lead to alternating regions of constructive and destructive interference, which manifest as enhanced or suppressed spectral features. In contrast, the quasiperiodic sequence largely avoids such extreme interference conditions, producing smoother and more distributed spectral profiles. In certain parameter regions, particularly when the periodic sequence is close to destructive interference, the quasiperiodic sequence can yield comparable or slightly larger particle production.

The particle yield reveals a more robust and less sensitive dependence on temporal ordering. For \(\xi_{12} \lesssim 1\), pair production remains strongly suppressed, and the yields obtained from periodic and quasiperiodic sequences are nearly identical. As \(\xi_{12}\) exceeds unity, the particle yield increases rapidly for both cases, reflecting the growing dominance of the stronger field component. Although modest differences between the two sequences are observed—typically with quasiperiodic driving yielding slightly higher values for \(\xi_{12} > 1\)—these differences remain within the same order of magnitude. This indicates that, while temporal ordering strongly affects the momentum distribution, the total particle yield is governed primarily by the effective field strength.

Increasing the pulse number from \(N = 12\) to \(N = 20\) enhances the visibility of interference effects in the momentum spectra, leading to sharper and more structured features. For periodic sequences, the interference peaks become narrower and more pronounced, consistent with coherent accumulation over longer pulse trains. For Fibonacci sequences, increasing \(N\) amplifies momentum-space localization and spectral fragmentation, indicating that quasiperiodicity-induced phase mismatch accumulates rather than averaging out.

Overall, our results demonstrate that engineering the temporal ordering of multi-pulse fields provides a versatile means of controlling vacuum pair production. Periodic temporal order enables strong and well-defined interference structures in momentum space, while quasiperiodic order offers a more uniform and robust spectral response. The distinction between these two regimes is particularly evident in the redistribution of spectral weight, even though the integrated yield remains comparatively insensitive to temporal ordering.

Future work could extend this analysis to other classes of aperiodic sequences, such as Thue–Morse or more general quasiperiodic modulations, to further explore the relationship between temporal complexity and quantum interference. Incorporating spatial inhomogeneities, magnetic fields, and fully three-dimensional pulse configurations would provide a more realistic description of experimentally relevant field setups. Such developments will be important for bridging the gap between theoretical predictions and the experimental realization of controlled vacuum pair production in next-generation high-intensity laser facilities.

\section*{Acknowledgments}

Deepak Sah acknowledges the financial assistance provided by the Raja Ramanna Center for Advanced Technology (RRCAT) and the Homi Bhabha National Institute (HBNI) for carrying out this research work.

%%%%%%%%%%%%%%%%%%%%%%%%%%%%%%%%%%%%%%%%%%%%%%%%%%%%%%%%%%%%%%%%%%%%%%%%%%%%%%%%%%%%%%%%%%%%%%%%%%%%%%%%%%%%%%%%%%%%%%%%%%%%%%%%%%%%%%%%%%%%%%%%%%%%%%%%%%%%%%%%%%%%%%%%%%%%%%%%%%%%%%%%%%%%%%%%%%%%%%%%%%%%%%%%%%%%%%%%%%%%%%%%%%%%%%%%%%%%%%%%%%%%%%%%%%%%%%%%%%%%%%%%%%%%%%%%%%%%%%%%%%%%%%%%%%%%%%%%%%%%%%%%%%%%%%%%%%%%%%%%%%%%%%%%%%%%%%%%%%%%%%%%%%%%%%%%%%%%%%%%%%%%%%%%%%%%%%%%%%%%%%%%%%%%%%%%%%%%%%%%%%%%%%%%%%%%%%%%%%%%%%%%%%%%%%%%%%%%%%%%%%%%%%%%%%%%%%%%%%%%%%%%%%%%%%%%%%%%%%%%%%%%%%%%%%%%%%%%%%%%%%%%%%%%%%%%%%%%%%%%%%%%%%%%%%%%%%%%%%%%%%%%%%%%%%%%%%%%%%%%%%%%%%%%%%%%%%%%%%%%%%%%%%%%%%%%%%%%%%%%%%%%%%%%%%%%%%%%%%%%%%%%%%%%%%%%%%%%%%%%%%%%%%%%%%%%%%%%%%%%%%%%%%%%%%%%%%%%%%%%%%%%%%%%%%%%%%%%%%%%%%%%%%%%%%%%%%%%%%%%%%%%%%%%%%%%%%%%%%%%%%%%%%%%%%%%%%%%%%%%%%%%%%%%%%%%%%%%%%%%%%%%%%%%%%%%%%%%%%%%%%%%%%%%%%%%%%%%%%%%%%%%%%%%%%%%%%%%%%%%%%%%%%%%%%%%%%%%%%%%%%%%%%%%%%%%%%%%%%%%%%%%%%%%%%%%%%%%%%%%%%%%%%%%%%%%%%%%%%%%%%%%%%%%%%%%%%%%%%%%%%%%%%%%%%%%%%%%%%%%%%%%%%%%%%%%%%%%
 %%%%%%%%%%%%%%%%%
\bibliographystyle{apsrev}
%\bibliography{ref}

\end{document}